\documentclass[a4paper,12pt]{article}

\usepackage{amssymb}
\usepackage{amsmath}
\usepackage{subeqnarray}
\usepackage{mathbbol}

\numberwithin{equation}{section}

\begin{document}

\title{Deconstruction of the Maldacena-N\'u\~nez Compactification}
\author{R.P. Andrews \footnote{pyrich@swan.ac.uk} \\ \\ Department of
  Physics, University of Wales Swansea \\ Singleton Park, 
Swansea, SA2 8PP, UK \\ \\ 
\and N. Dorey \footnote{N.Dorey@damtp.cam.ac.uk} \\ \\ Department of
Applied Mathematics and Theoretical Physics \\ 
Centre for Mathematical Sciences, University of Cambridge \\
Wilberforce Road, Cambridge, CB3 0WA, UK}
\date{}

\begin{figure}
\begin{flushright}
hep-th/0601098 \\
SWAT/06/455
\end{flushright}
\end{figure}

\maketitle

\begin{abstract}

We demonstrate a classical equivalence between the large-$N$ 
limit of the Higgsed 
$\mathcal{N}=1^*$ SUSY $U(N)$ Yang-Mills theory and 
the Maldacena-N\'u\~nez twisted compactification of a six
dimensional gauge theory on a two-sphere. 
A direct comparison of the actions and spectra of the two theories  
reveals them to be identical. We also propose a gauge theory limit
which should describe the corresponding spherical compactification of Little
String Theory.

\end{abstract}

\newpage

\section{Introduction}

An interesting development in string theory has been the discovery of
consistent interacting non-gravitational quantum theories in
dimensions greater than four\cite{bib:LST, bib:Aha}. 
A key example arising in six spacetime dimensions is Little String
Theory (LST). This theory 
emerges on  the world-volume of NS fivebranes
in critical string theory and decouples from gravity when the string
coupling is taken to zero. The resulting theory is 
non-local, exhibiting some string-like properties. At low energy
Little String Theory reduces to a conventional gauge
theory in six-dimensions. This low-energy gauge theory
is non-renormalisable and LST is believed to provide a consistent UV
completion.    
These theories are of great interest as they represent an
intermediate state between string theories and field theories. 
A greater understanding of little string theories could improve our 
understanding of string and gauge theories.
\paragraph{}
An important challenge in studying these higher-dimensional models is
to find a consistent non-perturbative definition of the theory. 
Deconstruction \cite{bib:AH}, \cite{bib:AHII}
\footnote{see also \cite{others}}  
is a promising approach to this problem 
which allows one to define higher-dimensional theories
 as certain limits of better understood four-dimensional gauge
theories. In particular, the idea is that a four-dimensional theory in its 
Higgs phase can be re-interpreted as a theory with extra compact,
discretized dimensions. It is hoped that in an appropriate continuum limit 
the higher-dimensional Lorentz invariance is restored. In recent work 
\cite{ND}, 
one of the authors proposed a new way of deconstructing a
toroidally-compactified LST using a marginal deformation of ${\cal
  N}=4$ SUSY Yang-Mills. One of the advantages of this approach is
that the continuum limit can be studied explicitly using S-duality and
the AdS/CFT correspondence. In this paper we will present some
preliminary evidence that a similar approach using a {\em relevant}
deformation of the ${\cal N}=4$ theory should give a deconstruction of the
spherical compactification of LST first studied by Maldacena and 
N\'u\~nez (MN) \cite{bib:MN}. 
\paragraph{}
The usual starting point for deconstruction is a four-dimensional
theory in its Higgs phase. In this case we will study the 
$\mathcal{N}=1^*$ SUSY Yang-Mills \cite{VW,Donagi, elliptic,bib:PS} 
with gauge group $U(N)$ in a vacuum
where the gauge group is broken down to a $U(1)$ subgroup. We will
show in detail that, at large-$N$, this theory is classically
equivalent to a twisted
compactification of ${\cal N}=(1,1)$ SUSY gauge theory in six
dimensions with gauge group $U(1)$. We demonstrate this equivalence 
by an explicit comparison of the classical spectra and the Lagrangians 
of the two theories in question, obtaining exact agreement. A
preliminary account of this work, including a comparison of the
fermionic spectra, appeared in our earlier letter \cite{bib:AD}. 
In the rest of this introductory section we give a brief overview of
these results and explain their potential relevance to the deconstruction of
LST.  
 \paragraph{}       
The ${\cal N}=1^{*}$ theory contains a $U(N)$
vector multiplet of ${\cal N}=1$ Supersymmetry and three adjoint
chiral multiplets of equal mass $\eta$. With appropriate rescaling the
mass parameter becomes an overall coefficient. Schematically, 
the superpotential is,
\begin{equation}
\mathcal{W}(\Phi) = \textrm{Tr}_N \left(  i \Phi_1 [\Phi_2, \Phi_3] + 
\frac{1}{2} \sum_{i=1}^{3} \Phi^2_i \right)
\end{equation}
which leads to the F-flatness condition, 
\begin{equation}
[\Phi_i, \Phi_j] = i\varepsilon_{ijk} \Phi_k
\end{equation}
which coincides with the Lie algebra of $SU(2)$. 
This can be solved by any $N$-dimensional representation of the
$SU(2)$ generators \cite{VW}. The same set of vacua are present in
both the $U(N)$ and $SU(N)$ cases. 
By choosing the vacuum $\langle\Phi_i\rangle = J^{(N)}_i$, the $N$-dimensional
representation of the $SU(2)$ generators, we break the gauge group
from $U(N)$ to $U(1)$ by the Higgs mechanism.
The emergence of the extra dimensions occurs via the mechanism seen 
in M(atrix) theory \cite{bib:Kimura}. The complex scalars of the
chiral multiplets form a fuzzy sphere: a discrete, non-commutative 
version of the 2-sphere. The expansion of the $\mathcal{N}=1^*$ fields 
about this vacuum allows the theory to be interpreted as a
six-dimensional non-commutative gauge theory with a UV cutoff in place for
finite $N$. Classically, in the limit $N \to \infty$, we get a
commutative, continuum theory on $\Re^{3,1} \times S^2$.
\paragraph{} 
The appearance of a six-dimensional theory can also be understood 
via the string theory realisation of the ${\cal N}=1^{*}$ theory \cite{bib:PS}.
The theory is realised on the worldvolume of $N$ D3 branes    
in the presence of a background three-form field strength leading 
to a version of the Myers effect, where the D3 branes polarize into 
a spherically wrapped D5 brane \cite{bib:Myers}. 
The theory living on the worldvolume 
of the D5 brane reduces to a six-dimensional $U(1)$ gauge
theory at low energies. The D3 brane charge is realised as $N$ units
of magnetic flux through the 2-sphere leading to non-commutativity
in the world-volume gauge theory \cite{bib:SW} as expected.  
The string theory picture also suggests the identity of the six-dimensional
theory. The worldvolume theory on a single D5 brane in flat space is a
$U(1)$ gauge theory 
in six dimensions with $\mathcal{N}=(1,1)$ supersymmetry. The 
string theory picture suggests that this world-volume theory should be
compactified on a 2-sphere in such a way that $\mathcal{N}=1$ supersymmetry is
preserved in the remaining four dimensions. A conventional compactification
on a 2-sphere would break all supersymmetries as there are 
no covariantly constant spinors on $S^2$. However, in \cite{bib:MN},
Maldacena and N\'u\~nez studied a particular 
twisted compactification preserving
a quarter of the supersymmetries. In their example, the
compactfication of the world-volume theory corresponded to wrapping 
the D5-brane on a non-contractible two-cycle of a Calabi-Yau
threefold. In the present case, the brane is wrapped on a
topologically trivial cycle supported by the presence of external
flux. Our main result is that, despite this difference, the resulting 
compactified world-volume theory is the same. In particular, as
$N\rightarrow\infty$, the
spectrum of the ${\cal N}=1^{*}$ theory reproduces that of the
six-dimensional theory. Interestingly, the agreement persists at
finite $N$ provided we make an appropriate truncation of the
six-dimensional Kaluza-Klein spectrum. 
We also check the agreement directly at the level of the Lagrangian.  
\paragraph{}
The MN compactification and the Higgsed $\mathcal{N}=1^*$ theory are
classically equivalent and are therefore different descriptions of the
same classical theory. Which description is most appropriate would
depend on the length scale probed. The compact extra dimensions have a
scale dictated by the radius of the 2-sphere $R$ which corresponds to 
inverse mass parameter $\eta^{-1}$ in the dual gauge theory. 
At distances $L> R$ the fields have insufficient 
energy to probe the extra dimensions
so the theory appears to be a four-dimensional 
$\mathcal{N}=1$ theory with gauge group $U(1)$ and coupling $G^2_4 =
g^2_{ym}/N$. At distances $L < R$ the fields have sufficient
energy to propagate in the extra dimensions and the theory appears to
be the six-dimensional MN compactification with gauge group $U(1)$ and 
coupling $G^2_6 = 4 \pi R^2 g^2_{ym}/N$ . As mentioned above the
Kaluza-Klein spectrum is truncated at finite $N$. Correspondingly, at 
distances $L < R/N$, the theory becomes
four-dimensional again and the full $U(N)$ gauge symmetry is
restored. Thus the length-scale $R/N\sim 1/(N\eta)$ is
the effective UV cutoff or lattice spacing of the six dimensional
theory. This behaviour is typical of deconstruction.

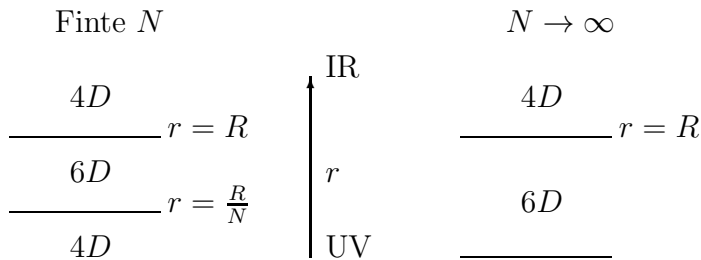
\begin{flushleft}
\begin{picture}(20,100)
\setlength{\unitlength}{2cm}
\put(1,0.3){\line(1,0){1}}
\put(1,0.8){\line(1,0){1}}
\put(3,0){\vector(0,1){1.2}}
\put(3.1,0.5){$r$}
\put(1.4,0){$4D$}
\put(1.4,1){$4D$}
\put(1.4,0.5){$6D$}
\put(2.05,0.3){$r=\frac{R}{N}$}
\put(2.05,0.8){$r=R$}
\put(1.3,1.5){Finte $N$}
\put(3.1,0){UV}
\put(3.1,1.2){IR}
\put(4,0){\line(1,0){1}}
\put(4,0.8){\line(1,0){1}}
\put(4.4,1){$4D$}
\put(4.4,0.3){$6D$}
\put(5.05,0.8){$r=R$}
\put(4.3,1.5){$N \to \infty$}
\end{picture}
\end{flushleft}

An obvious question is whether we can use the four-dimensional gauge
theory to define a continuum limit of the six-dimensional theory where
the lattice spacing goes to zero. If so we would also like to identify
the resulting continuum theory. The situation is closely related to
the one described in detail in \cite{ND} and we can give a very
similar discussion in the present case. 
To get an interacting theory we must 
keep the six-dimensional gauge coupling $G_{6}$ fixed. It will
also be convenient to keep the compactification radius $R$ fixed. The
corresponding gauge theory limit involves taking $N\rightarrow \infty$ 
with $\eta$ and $g^{2}_{ym}/N$ held fixed. Thus we see that the continuum
limit of the six-dimensional theory is a strong coupling limit of the
four-dimensional gauge theory. This means we cannot rely on 
a classical analysis of either theory to
study the continuum limit. This again is typical of deconstruction.     
\paragraph{}
However, in the present context, we can make some progress using the
exact S-duality of the ${\cal N}=1^{*}$ theory which takes
$g^{2}_{ym}\rightarrow \tilde{g}^{2}_{ym}=16\pi^{2}/g^{2}_{ym}$ and 
maps the Higgs phase vacuum
studied above to a vacuum in a confining phase \cite{Donagi,elliptic}. 
The mass parameter
$\eta$ is invariant under S-duality. Under this transformation our
proposed continuum limit becomes a 't Hooft large $N$ limit where
$N\rightarrow \infty$ with $\tilde{g}^{2}_{ym}N$ and $\eta$ are held fixed. 
As usual, we expect that this limit makes sense and leads to a theory
with the qualitative features of a closed string theory. 
Indeed for $\tilde{g}^{2}_{ym}N>>1$, the corresponding $SU(N)$ 
theory in this vacuum has a dual
description involving a spherically wrapped NS5 brane in an
asymptotically AdS geometry \cite{bib:PS}. 
The proposed continuum limit corresponds to a
limit of this system where the dual string coupling 
$g_{s}=\tilde{g}^{2}_{ym}/4\pi\rightarrow 0$.  
\paragraph{}
In the case of multiple coincident
NS-five branes this is precisely the limit used to define LST. 
Unfortunately, in the present case of a single NS-fivebrane it is not
known whether a decoupled world-volume theory exists. However the
above results immediately suggest a promising generalisation which involves
multiple NS fivebranes. Namely, we should start from an $SU(N)$ theory in the
vacuum where the $\Phi_{i}$ take VEVs corresponding to the sum of 
$p$ irreducible representations of $SU(2)$ of dimension $q$; 
\begin{equation*}
<\Phi_i> = \mathbb{1}_{(p)} \otimes J^{(q)}_i
\end{equation*}
This corresponds to an $SU(N)$ theory with $N=pq$ Higgsed down to $SU(p)$. 
The low-energy effective theory is ${\cal N}=1$ SUSY Yang-Mills with
gauge group $SU(p)$ which confines in the IR at scales of order 
$\Lambda\sim \eta\exp(-8\pi^{2}q/g^{2}_{ym}p)$. However, at scales far
above $\Lambda$, the theory is weakly coupled and 
we may identify the effective theory by studying the classical action. 
Indeed a straightforward generalisation of the above results, discussed in
Section 5, suggests that the effective theory is a twisted
compactification of a six-dimensional $SU(p)$ gauge theory of the 
Maldacena-Nunez type. Taking a
similar continuum limit, $q\rightarrow \infty$ with $p$, $g^{2}_{ym}/q$
and $\eta$ held fixed we may perform S-duality to a phase where $U(N)$ is
Higgsed to $U(q)$ which is then confined at a lower
scale. As before the corresponding limit becomes a 't Hooft-like limit
where $N=pq\rightarrow \infty$ with $\tilde{g}^{2}_{ym}q$, $p$ and $\eta$ 
held fixed. The IIB dual
now involves $p$ spherically wrapped NS fivebranes \cite{bib:PS} 
in a decoupling limit where $g_{s}\rightarrow 0$. 
This suggests the ${\cal N}=1^*$ theory really
deconstructs the full Little String Theory on the fivebrane
worldvolume and not just the low energy
six-dimensional gauge theory. More precisely, as in the toroidal case
of \cite{ND} and in \cite{BD},
a sub-sector of the four-dimensional theory retains non-vanishing
interactions in this limit while the remaining states in the theory
decouple. We propose that the interacting sector is
equivalent to the MN compactification of LST in this
limit. The Little Strings themselves correspond to the confining chromoelectric
flux tube of the gauge theory which has fixed tension in the 't Hooft
limit.            
\paragraph{}
The rest of the paper is organised as follows. In section 2 we discuss
scalars, spinors and vectors on the 2-sphere. In section 3, we determine 
the Kaluza-Klein
spectrum of the MN compactification and the corresponding
action for $\mathcal{N}=(1,1)$ SUSY
Yang-Mills theory on $\mathcal{R}^{3,1}\times S^{2}$. 
In section 4 we describe
the emergence of extra dimensions in the Higgsed $\mathcal{N}=1^*$
theory via deconstruction.
In section 5 we review the calculation of the
fermionic spectrum of the $\mathcal{N}=1^*$ theory from \cite{bib:AD}
in greater detail (for a related calculation see \cite{Por}) 
and calculate the associated bosonic spectrum. We
then compare the $\mathcal{N}=1^*$ spectrum with the Kaluza-Klein
spectrum of the MN compactification and show them to be identical in
the limit $N \to \infty$. We also discuss briefly the generalisations of our
result for other ${\cal N}=1^*$ vacua. 
In section 6 we will calculate the effective
six-dimensional action of the Higgsed $\mathcal{N}=1^*$ theory. We see
that the action is identical to the action of section \ref{sec:MNA},
demonstrating the equivalence of the MN compactification and the
Higgsed $\mathcal{N}=1^*$ theory. In Appendix \ref{app:1} we 
discuss Clifford algebras in various dimensions and
review the dimensional reduction of 
$\mathcal{N}=1$ SUSY Yang-Mills theory in ten spacetime dimensions to
six spacetime dimensions obtaining the $\mathcal{N}=(1,1)$ SUSY Yang-Mills 
theory.

\section{Eigenvalues and Eigenstates on the 2-Sphere}\label{sec:sphere}

In this section we will discuss fields of different spin on the
2-sphere. To fix our notation and conventions we begin with scalar
fields before describing the less well-known subjects of spinors 
and vectors on the 2-sphere. The 2-sphere is a two-dimensional manifold 
embedded in $\Re^3$, with a global $SO(3) \sim SU(2)$ isometry group, defined 
by the equation,
\begin{equation}
x^2_1 + x^2_2 + x^2_3 = R^2
\end{equation}
for a coordinate basis $x_i$ in $\Re^3$. We define the coordinates $x_i$ in 
terms of the coordinates on the 2-sphere $q_a = (\theta, \phi)$ and radius $R$ 
by,
\begin{subeqnarray}
x_1 & = & R \, \textrm{sin} \, \theta \, \textrm{cos} \, \phi \\
x_2 & = & R \, \textrm{sin} \, \theta \, \textrm{sin} \, \phi \\
x_3 & = & R \, \textrm{cos} \, \theta
\end{subeqnarray}
which dictates the metric of the 2-sphere,
\begin{equation}
ds^2 = R^2 \, d\theta^2 + R^2 \, \textrm{sin}^2 \theta \, d\phi^2
\end{equation}
The generators of $SU(2) \sim SO(3)$ are the angular momentum operators $L_i$.
\begin{equation}
L_i = -i \varepsilon_{ijk} x_j \partial_k 
\end{equation}
In terms of coordinates on the 2-sphere the angular momentum operators are,
\begin{subeqnarray}
L_1 & = & \phantom{-} i \, \textrm{sin} \, \phi \, \frac{\partial}{\partial 
\theta} + i \, \textrm{cos} \, \phi \, \textrm{cot} \, \theta \, 
\frac{\partial}{\partial \phi} \\
L_2 & = & -i \, \textrm{cos} \, \phi \, \frac{\partial}{\partial \theta} + 
i\, \textrm{sin} \, \phi \, \textrm{cot} \, \theta \, 
\frac{\partial}{\partial \phi} \\
L_3 & = & -i \, \frac{\partial}{\partial \phi}
\end{subeqnarray}
which we can summarize as \cite{bib:Kimura},
\begin{equation}\label{eq:killing}
L_i = -i k^a_i \partial_a
\end{equation}
The metric tensor can also be expressed in terms of the Killing
vectors $k^a_i$ (defined by the above equations) as,
\begin{equation}
g^{ab} = \frac{1}{R^2} \, k^a_i k^b_i
\end{equation}

We can expand any function on the 2-sphere in terms of the eigenfunctions of 
the 2-sphere,
\begin{equation}\label{eq:expand}
a(\theta, \phi) = \sum^{\infty}_{l=0} \sum^l_{m=-l} a_{lm}
Y_{lm}(\theta, \phi) 
\end{equation}
where $a_{lm}$ is a complex coefficient 
and $Y_{lm}(\theta, \phi)$ are the 
spherical harmonics, which satisfy the equation,
\begin{equation}
L^2 Y_{lm} = -R^2 \Delta_{S^2} Y_{lm} = l(l+1) Y_{lm}
\end{equation}
where $\Delta_{S^2}$ is the scalar Laplacian on the 2-sphere,
\begin{equation}
\Delta_{S^2} = \frac{1}{\sqrt{g}} \, \partial_a ( g^{ab} \sqrt{g} \, \partial_b)
\end{equation}
The spherical harmonics have an eigenvalue $\mu \sim l(l+1)$ for integer \mbox{$l = 0,1,
\dots$}, with degeneracy $2l+1$. The orthogonality condition of the spherical 
harmonics is,
\begin{equation}
\int d\Omega \, Y^{\dag}_{lm} Y^{\phantom{\dag}}_{l'm'} = \delta_{l l'} \, 
\delta_{m m'}
\end{equation}
where $d\Omega = \sin \theta \, d\theta d\phi$.
\paragraph{}
The eigenstates of spin-$\frac{1}{2}$ particles on the 2-sphere are the 
spherical spinors. They are eigenstates of the total angular momentum, 
$\hat{L}^2$ and $\hat{L}_3$. We form spherical spinors from spherical harmonics
and the spin-$\frac{1}{2}$ eigenstates of spin operators $S^2$ and $S_3$,
\begin{equation}\label{eq:spin1/2}
\chi_{\frac{1}{2}} = \left(\begin{array}{c} 1 \\ 0 \end{array}\right) \quad 
\chi_{-\frac{1}{2}} =  \left(\begin{array}{c} 0 \\ 1 \end{array}\right)
\end{equation}
The spherical spinors are \cite{bib:Atomic},
\begin{equation}
\Omega_{jlm}(\theta, \phi) = \sum_{\mu} C(l,\frac{1}{2},j; m-\mu, \mu, m) 
Y_{l, m-\mu}(\theta, \phi) \chi_{\mu}
\end{equation}
where $C(l,\frac{1}{2},j; m-\mu, \mu, m)$ are Clebsch-Gordan coefficients.
Explicitly the spherical spinors are,
\begin{eqnarray}
\Omega^{\hat{\alpha}}_{l + \frac{1}{2},lm}(\theta, \phi) & = & \left( 
\begin{array}{c} \sqrt{\frac{l+m+\frac{1}{2}}{2l+1}} \, Y_{l, m-\frac{1}{2}}
(\theta, \phi) \\ \sqrt{\frac{l-m+\frac{1}{2}}{2l+1}} \, Y_{l, m+\frac{1}{2}}
(\theta, \phi) \end{array} \right) \\
\Omega^{\hat{\alpha}}_{l - \frac{1}{2},lm}(\theta, \phi) & = & \left( 
\begin{array}{c} -\sqrt{\frac{l-m+\frac{1}{2}}{2l+1}} \, Y_{l, m-\frac{1}{2}}
(\theta, \phi) \\ \sqrt{\frac{l+m+\frac{1}{2}}{2l+1}} \, Y_{l, m+\frac{1}{2}}
(\theta, \phi) \end{array} \right)
\end{eqnarray}
where $\hat{\alpha}$ is a spinor index labelling the two components.
The spherical spinors are eigenstates of the Dirac operator on the 2-sphere,
\begin{equation}
\kappa = -\frac{1}{R} \, (1 + \sigma_i L_i)
\end{equation}
Here, the index $i$ on $L_i$ refers to the 
cartesian basis of the
 three dimensional Euclidean embedding space of the two sphere. 
The operator has eigenvalues,
\begin{equation}
\kappa^{\hat{\alpha}}_{\phantom{\alpha} \hat{\beta}} \, \Omega^{\hat{\beta}}_{q_{\pm}lm}(\theta, \phi) = \frac{1}{R} \, \kappa_{\pm} \, \Omega^{\hat{\alpha}}_{q_{\pm}lm}(\theta, \phi)
\end{equation}
where $\kappa_{\pm} = \mp (q_{\pm} + \frac{1}{2}), \ q_{\pm} = l \pm 
\frac{1}{2}$. 
In analogy to the spherical harmonics we can expand a spinor on the 2-sphere in
terms of the spherical spinors,
\begin{equation}
\psi^{\hat{\alpha}}(\theta, \phi) = \sum^{\infty}_{l=0} 
\sum^{q_{\pm}}_{m=-q_{\pm}} \left\lbrace \psi^{(+)}_{lm} 
\Omega^{\hat{\alpha}}_{l + \frac{1}{2},lm}(\theta, \phi) + \psi^{(-)}_{lm} \Omega^{\hat{\alpha}}_{l - \frac{1}{2},lm}(\theta, \phi) \right\rbrace
\end{equation}
where $\psi^{(\pm)}_{lm}$ are complex coefficients.
The orthogonality condition of the spherical spinors is inherited from the 
spherical harmonics,
\begin{equation}\label{eq:sphorth}
\int d\Omega \; \Omega^{\dag}_{q m \hat{\alpha}}(\theta, \phi) \, 
\Omega^{\hat{\alpha}}_{q' m'}(\theta, \phi) = \delta_{q q'} \delta_{m m'}
\end{equation}

The spherical spinors can also be represented in a spherical basis, in terms of
 coordinates on the 2-sphere $q^a$ \cite{bib:Abri}, which takes advantage of 
 the local $SO(2)$ symmetry of the 2-sphere.
The zweibein (and metric tensor) on the unit 2-sphere are,
\begin{equation}
g_{ab} = \textrm{diag}(R^2, R^2 \, \textrm{sin}^2 \theta); \qquad e^{\alpha}_a = 
\textrm{diag}(R, R \, \textrm{sin} \, \theta); \qquad g_{ab} = \delta_{\alpha \beta}
 \, e^{\alpha}_a \, e^{\beta}_b
\end{equation}
where we denote the coordinates on the 2-sphere by the index $a,b$ and the 
local frame by $\alpha,\beta$. Derivatives on the 2-sphere are replaced by 
generally covariant derivatives,
\begin{equation}
\nabla_{a} \Upsilon = \partial_{a} \Upsilon + \frac{i}{4} \, R^{\alpha \beta}_{a} 
\sigma_{\alpha \beta} \Upsilon
\end{equation}
where $R^{\alpha \beta}_{a}$ is the spin connection on the 2-sphere and 
$\sigma_{\alpha \beta}$ are spin-$\frac{1}{2}$ generators of $SO(2)$. The 
non-zero components of the spin connection on the 2-sphere are,
\begin{equation}
R^{12}_{\phi} = -R^{21}_{\phi} = - \textrm{cos} \, \theta
\end{equation}
The Clifford algebra in two dimensions is $\hat{\gamma}^a = \lbrace \sigma_1, 
\sigma_2 \rbrace$, where $\sigma_i$ are the Pauli matrices. We define the 
generators $\sigma_{\alpha \beta}$ through the Clifford algebra,
\begin{equation}
\sigma_{12} = - \sigma_{21} = -\frac{i}{2} [\hat{\gamma}_1, \hat{\gamma}_2] = \sigma_3
\end{equation}
The Dirac operator on the 2-sphere is defined as,
\begin{equation}
-i\hat{\nabla}_{S^2} = -i e^{a \alpha} \sigma_{\alpha} \nabla_a
\end{equation}
Explicitly the Dirac operator is,
\begin{equation}
-i \hat{\nabla}_{S^2} = -\frac{i \sigma_1}{R} \left(\frac{\partial}{\partial \theta} + 
\frac{\textrm{cot} \theta}{2} \right) - \frac{i \sigma_2}{R \, \textrm{sin} \, 
\theta} \frac{\partial}{\partial \phi}
\end{equation}

We calculate the eigenvalues of this Dirac operator by considering the squared
Dirac operator,
\begin{equation}\begin{split}
(-i \hat{\nabla}_{S^2})^2 = -\frac{1}{R^2} \bigg( \textrm{cot} & \, \theta \, \partial_{\theta} + 
\partial_{\theta} \partial_{\theta} + \textrm{csc}^2 \theta \, \partial_{\phi} 
\partial_{\phi} \\
& - i \sigma_3 \, \textrm{csc} \, \theta \, \textrm{cot} \, 
\theta \, \partial_{\phi} - \frac{1}{4} - \frac{1}{4} \textrm{csc}^2 \theta \bigg)
\end{split}
\end{equation}
The generators of $SU(2)$ in the spin-$\frac{1}{2}$ representation are,
\begin{subeqnarray}
\hat{L}_3 & = & -i \partial_{\phi} \\
\hat{L}_{\pm} & = & \pm \, e^{\pm i \phi} \left( \partial_{\theta} \pm i 
\textrm{cot} \, \theta \, \partial_{\phi} \pm \frac{1}{2} \, \textrm{sin} \, 
\theta \, \sigma_3 \right)
\end{subeqnarray}
We find the relation between the Casmir operator $\hat{L}^2$ and the square of 
the Dirac operator is,
\begin{equation}
R^2 \, (-i \hat{\nabla}_{S^2})^2 = \hat{L}^2 + \frac{1}{4}
\end{equation}
Therefore the eigenvalues are related and the eigenvalue equation with 
spherical spinor $\Upsilon$ is,
\begin{equation}
\begin{split}
\left(-i \hat{\nabla}_{S^2}\right)^{2} \Upsilon_{jm} & = \frac{1}{R^2} \, \left( 
\hat{L}^2 + \frac{1}{4} \right) \Upsilon_{jm} \\
& = \frac{1}{R^2} \, \left( j(j+1) + \frac{1}{4} \right) \Upsilon_{jm}
\end{split}
\end{equation}
the eigenvalue is $\mu \sim j(j+1) + \frac{1}{4}$ for $j = \frac{1}{2}, 
\frac{3}{2}, \dots$, with degeneracy $2j+1$. By setting $j = \frac{2l-1}{2}$, 
$l = 1,2,\dots$ we re-express the eigenvalues in terms of an integer quantum 
number $l$, resulting in $\mu \sim l^2$. The allowed eigenvalues correspond to 
$\mu \sim l^{2}$ for integer $l\geq 1$ with degeneracy $2l$ \cite{bib:Abri}. 

We have two types of spinor, both are a complete orthonormal set of spinors on 
the 2-sphere. The two types of the spherical spinors, cartesian $\Omega$ and 
spherical $\Upsilon$ are related via a spinor transformation \cite{bib:Abri}. 
If we consider two spinors, one in the cartesian basis $\psi(x)$ and one in the
spherical basis $\psi(q)$. The transformation is,
\begin{equation}\label{eq:basetran}
\psi^{\hat{\alpha}}(x) = (V^{\dag})^{\hat{\alpha}}_{\phantom{\alpha} 
\hat{\beta}} \psi^{\hat{\beta}}(q)
\end{equation}
where the matrix $V$ is,
\begin{equation}
V =  \left(\begin{array}{cc} e^{i\frac{\phi}{2}} \textrm{cos} \, 
\left(\frac{\theta}{2}\right) & e^{-i\frac{\phi}{2}} \textrm{sin} \, \left(\frac{\theta}{2}\right) \\ -e^{i\frac{\phi}{2}} \textrm{sin} \, \left(\frac{\theta}{2}\right) & e^{-i\frac{\phi}{2}} \textrm{cos} \, \left(\frac{\theta}{2}\right) \end{array}\right)
\end{equation}
The Dirac operators are related by the similarity transformation,
\begin{equation}\label{eq:basetran2}
\left(\hat{\gamma}_3\right)^{\hat{\alpha}}_{\phantom{\alpha} \hat{\gamma}}(-i \hat{\nabla}_{S^2})^{\hat{\gamma}}_{\phantom{\alpha} \hat{\beta}} = V^{\hat{\alpha}}_{\phantom{\alpha} \hat{\gamma}} \; 
\kappa^{\hat{\gamma}}_{\phantom{\gamma} \hat{\delta}} \, (V^{\dag})^{\hat{\delta}}_{\phantom{\delta} \hat{\beta}}
\end{equation}
where $\hat{\gamma}_3 = \sigma_1 \sigma_2$ is the two-dimensional chirality operator.
The spherical spinors types diagonalize two operators \cite{bib:Abri}, both 
diagonalize the total angular momentum $\hat{L}^2_i$ (by definition), the 
spherical spinor $\Upsilon$ diagonalizes the Dirac operator on the 2-sphere $(-i\hat{\nabla}_{S^2})^2$ whilst the spherical spinor $\Omega$ diagonalizes 
the orbital angular momentum $L^2_i$.
\paragraph{}
With vector fields on the 2-sphere, it is not consistent to expand the 
components of the vector separately in spherical harmonics \cite{bib:Mag}, 
e.g.
\begin{equation}
\begin{split}
{\bf V}(r, \theta, \phi) & = \hat{{\bf e}}_{r} V^r + 
\hat{{\bf e}}_{\theta} V^{\theta} + \hat{{\bf e}}_{\phi} V^{\phi} \\
& = \hat{{\bf e}}_{r} \sum_{jm} v^{r}_{jm} \, Y_{jm} + 
\hat{{\bf e}}_{\theta} \sum_{jm} v^{\theta}_{jm} \, Y_{jm} + 
\hat{{\bf e}}_{\phi} \sum_{jm} v^{\phi}_{jm} \, Y_{jm}
\end{split}
\end{equation}
We must form eigenstates of the total angular momentum for spin-$1$ particles,
the vector harmonics are formed from spherical harmonics and the spin-$1$ 
eigenstates of $S^2$ and $S_3$,
\begin{equation}\label{eq:spin1}
\xi_1 = -\frac{1}{\sqrt{2}}\left(\begin{array}{c} 1 \\ i \\ 0 \end{array}
\right) \quad \xi_0 =  \left(\begin{array}{c} 0 \\ 0 \\ 1 \end{array}\right)
\quad \xi_{-1} = \frac{1}{\sqrt{2}} \left(\begin{array}{c} 1 \\ -i \\ 0 
\end{array}\right) 
\end{equation}
The vector harmonics are \cite{bib:Atomic},
\begin{equation}
{\bf Y}_{jlm}(\theta, \phi) = \sum_{\sigma} C(l,1,j; m-\sigma,\sigma,m) 
Y_{l,m-\sigma}(\theta, \phi) \xi_{\sigma}
\end{equation}
where $j,l,m$ are integer quantum numbers. 
We will use a basis which takes advantage of the $SO(2)$ `Lorentz' symmetry 
\cite{bib:Atomic, bib:Mag}, 
\begin{subeqnarray}
\tilde{{\bf T}}_{jm} & = & {\bf Y}_{jjm}(\theta,\phi) = \frac{1}{\sqrt{j(j+1)}} \left[\frac{\partial Y_{jm}}{\partial \theta} \, \mbox{\boldmath $\hat{\phi}$} 
- \textrm{csc} \, \theta \frac{\partial Y_{jm}}{\partial \phi} \, 
\mbox{\boldmath $\hat{\theta}$} \right] \\
\nonumber & = & \frac{1}{\sqrt{j(j+1)}} \, i \, {\bf L} \, Y_{jm} \\
\tilde{{\bf S}}_{jm} & = & \sqrt{\frac{j+1}{2j+1}} {\bf Y}_{j j-1 m}(\theta, 
\phi) + \sqrt{\frac{j}{2j+1}} {\bf Y}_{j j+1 m}(\theta, \phi) \\
\nonumber & = & \frac{1}{\sqrt{j(j+1)}} \left[\frac{\partial Y_{jm}}{\partial 
\theta} \, \mbox{\boldmath $\hat{\theta}$} + \textrm{csc} \, \theta 
\frac{\partial Y_{jm}}{\partial \phi} \, \mbox{\boldmath $\hat{\phi}$} \right] 
= \frac{1}{\sqrt{j(j+1)}} \, \mbox{\boldmath $\partial$} \, Y_{jm} \\
\tilde{{\bf R}}_{jm} & = & \sqrt{\frac{j}{2j+1}} {\bf Y}_{j j-1 m}
(\theta, \phi) - \sqrt{\frac{j+1}{2j+1}} {\bf Y}_{j j+1 m}(\theta, \phi) \\
\nonumber & = & \hat{\bf r} \, Y_{jm}
\end{subeqnarray}
The harmonics $\tilde{{\bf T}}_{jm}$ and $\tilde{{\bf S}}_{jm}$ are tangential 
to the 2-sphere, $\tilde{{\bf R}}_{jm}$ is normal to the 2-sphere.
Restricting to the vectors on a unit 2-sphere, we set the radial unit vector 
$\hat{\bf r}=0$, therefore $\tilde{{\bf R}}_{jm} =0$.
Covariant and contravariant vectors on the 2-sphere can then be
defined as, 
\begin{equation}
\tilde{V}_i = h_i V^i = h^{-1}_i V_i
\end{equation}
where $g_{ij} = h^2_i \delta_{ij}$. The corresponding 
covariant vector harmonics are,
\begin{subeqnarray}
\frac{1}{R} \, {\bf T}_{jm} & = & \frac{1}{\sqrt{j(j+1)}} \, \left[\textrm{sin} \, \theta \, 
\partial_{\theta} Y_{jm} \, \mbox{\boldmath $\hat{\phi}$} - \textrm{csc} \, 
\theta \, \partial_{\phi} Y_{jm} \, \mbox{\boldmath $\hat{\theta}$} \right] \\
\frac{1}{R} \, {\bf S}_{jm} & = & \frac{1}{\sqrt{j(j+1)}} \, \left[\partial_{\theta} Y_{jm} \, 
\mbox{\boldmath $\hat{\theta}$} + \partial_{\phi} Y_{jm} \, 
\mbox{\boldmath $\hat{\phi}$} \right]
\end{subeqnarray}

The Maxwell field on a 2-sphere is a vector field with the gauge
invariance, 
\begin{equation}
A_{\mu} \to A_{\mu} - R \, \partial_{\mu} \chi
\end{equation}
The gauge fields can be expanded in vector harmonics and the scalar
$\chi$ can be expanded in spherical harmonics. Under a gauge 
transformation the components transform as, 
\begin{eqnarray}
A'_{\theta}(\theta, \phi) & = & R \, \sum_{jm} \left( t_{jm} \, (-\textrm{csc}
\, \theta) \, \partial_{\phi} Y_{jm} + s_{jm} \partial_{\theta} Y_{jm}
- \chi_{jm} \partial_{\theta} Y_{jm} \right)  \nonumber \\
A'_{\phi}(\theta, \phi) & = & R \, \sum_{jm} \left( t_{jm} \, \textrm{sin}
\, \theta \, \partial_{\theta} Y_{jm} + s_{jm} \, \partial_{\phi} Y_{jm} -
\chi_{jm} \, \partial_{\phi} Y_{jm} \right) \nonumber
\end{eqnarray}
It therefore follows that we can set the complex coefficient $s_{jm} = 0$ via
a gauge transformation with $\chi_{jm}=s_{jm}$. The corresponding 
gauge fixing condition is the generally covariant analogue of the 
Lorentz Gauge.
\begin{eqnarray}
\nonumber \nabla^a A_a & = & g^{ab} \nabla_b A_a \\
& = & g^{ab} \partial_b A_a - g^{ab} \Gamma^c_{ba} A_c
\end{eqnarray}
For the 2-sphere there are only three non-zero Christoffel symbols,
\begin{equation}
\Gamma^{\theta}_{\phi \phi} = - \textrm{cos} \, \theta \, \textrm{sin} \, 
\theta \qquad \Gamma^{\phi}_{\theta \phi} = \Gamma^{\phi}_{\phi \theta} = 
\textrm{cot} \, \theta
\end{equation}
Therefore,
\begin{equation}
\begin{split}
\nabla^a A_a & = g^{ab} \partial_b A_a + \frac{1}{R^2} \, \textrm{cot} \, \theta 
\, A_{\theta} \\
& = \frac{1}{R} \, \sum_{jm} \frac{1}{\sqrt{j(j+1)}} \bigg\lbrace t_{jm} \Big( 
\partial_{\theta} (-\textrm{csc} \, \theta \, \partial_{\phi} Y_{jm}) - 
\textrm{cot} \, \theta \textrm{csc} \, \theta \, \partial_{\phi} Y_{jm} \\
& \quad + \textrm{csc} \, \theta \, \partial_{\phi} \partial_{\theta} 
Y_{jm} \Big)  + s_{jm} \Big( \partial_{\theta} \partial_{\theta} Y_{jm} + 
\textrm{cot} \, \theta \, \partial_{\theta} Y_{jm} + \textrm{csc}^2 \theta 
\, \partial_{\phi} \partial_{\phi} Y_{jm} \Big) \bigg\rbrace \\
& = -R \, \sum_{jm} \frac{1}{\sqrt{j(j+1)}} \, s_{jm} \Delta_{S^2} Y_{jm}
\end{split}
\end{equation}
The gauge condition $\nabla^a A_a = 0$ thus sets $s_{jm} = 0$.
The orthonormality condition for the vector harmonic ${\bf T}_{lm}$ is,
\begin{equation}
\int d\Omega \, {\bf T}^{\dag}_{lm} {\bf T}^{\phantom{\dag}}_{l'm'} = 
\delta_{ll'} \delta_{mm'}
\end{equation}
And it is an eigenfunction of the total angular momentum $\hat{L}^2$ and orbital angular momentum $L^2$,
\begin{equation}
L^2 {\bf T}_{lm} = l(l+1) {\bf T}_{lm}
\end{equation}
It follows that we can expand any gauge field on the 2-sphere as,
\begin{equation}
{\bf A} = \sum^{\infty}_{l=1} \sum^l_{m=-l} a_{lm} {\bf T_{lm}} 
\end{equation}

\section{Twisted Compactification}\label{sec:MN}

In this Section we study the 
Maldacena-N\'u\~nez compactification of $\mathcal{N}=(1,1)$ SUSY 
Yang-Mills in six dimensions and 
its classical spectrum of Kaluza-Klein modes as in \cite{bib:AD}. 
We start from the $U(1)$ theory defined on a six-dimensional Minkowski 
space $\Re^{5,1}$. The global symmetry group is
\begin{equation}
SO(5,1) \times SO(4) \simeq SU(4) \times SU(2)_A \times SU(2)_B 
\end{equation}
Here $SO(5,1)$ is the six-dimensional Lorentz group and 
$SO(4)$ is the R-symmetry of the ${\cal N}=(1,1)$ superalgebra. 
The matter content is a single $U(1)$ vector multiplet of 
$\mathcal{N} = (1,1)$ supersymmetry. It contains
\footnote{The corresponding spacetime indices run over
$M=0,1,\dots, 5$, $i=1,\dots,4$, $l=1,\dots,4$, $\bar{l}=1,\dots,4$} 
a six-dimensional gauge
field $A_M$, four real scalar fields $\phi_{i}$ and two 
Weyl spinors of opposite chirality; $\lambda_{l}$ and 
$\tilde{\lambda}_{\bar{l}}$. 
Their transformation properties under the global symmetries are,
\begin{center}
\begin{tabular}{l c c c}
\hline
& $SU(4)$ & $SU(2)_A$ & $SU(2)_B$ \\
\hline
$A_M$ & ${\bf 6}$ & ${\bf 1}$ & ${\bf 1}$ \\
$\phi_i$ & ${\bf 1}$ & ${\bf 2}$ & ${\bf 2}$ \\
$\lambda_l$ & ${\bf 4}$ & ${\bf 2}$ & ${\bf 1}$\\
$\tilde{\lambda}_{\bar{l}}$ & ${\bf \bar{4}}$ & ${\bf 1}$ & ${\bf 2}$ \\
\hline
\end{tabular}
\end{center}
\paragraph{}
Anticipating compactification of two spatial dimensions, we write the
space-time as $\Re^{5,1} \sim \Re^{3,1} \times \Re^2$. 
This decomposition breaks the six-dimensional Lorentz group down to 
a subgroup, 
\begin{equation}
H=SO(3,1) \times SO(2) 
\end{equation}
with covering group, 
\begin{equation}
\bar{H}=SU(2)_L \times SU(2)_R \times U(1)_{45}
\end{equation}
It is straightforward to decompose 
the six-dimensional fields into representations of $H$. 
The gauge field is written as, 
\begin{eqnarray}
A_M & = & A_{\mu}  \qquad \mu = 0,1,2,3\\
\nonumber & = & A_a \qquad \ a = M-3 = 4,5
\end{eqnarray}
and we define the complex fields (taking $SO(2) \to U(1)$),
\begin{equation}
n_{\pm} = \frac{1}{\sqrt{2}} \left( A_4 \pm i A_5 \right)
\end{equation}
Under the decomposition of the $SU(4)$ covering group,  
the six-dimensional spinors $\lambda_{l}$ and $\tilde{\lambda}_{\bar{l}}$,
transforming in the 
${\bf 4}$ and $\bar{\bf 4}$, split according to, 
\begin{eqnarray}
{\bf 4}  & \rightarrow & ({\bf 2},{\bf 1})^{+1}\oplus 
({\bf 1},{\bf 2})^{-1} \nonumber \\ 
 \bar{\bf 4}  & \rightarrow & ({\bf 2},{\bf 1})^{-1}\oplus 
({\bf 1},{\bf 2})^{+1}  \nonumber
\end{eqnarray}
under $SU(2)_{L}\times SU(2)_{R}$ where the superscript denotes 
$U(1)_{45}$ charge. 
Thus we obtain a total of four left-handed Weyl spinors 
$\lambda^{\alpha}_{\underline{\alpha}}$,
$\psi^{\alpha}_{\underline{\dot{\alpha}}}$ and four right-handed spinors 
$\bar{\lambda}^{\dot{\alpha}}_{\underline{\alpha}}$,
$\bar{\psi}^{\dot{\alpha}}_{\underline{\dot{\alpha}}}$. Here $\alpha$ and 
$\dot{\alpha}$ are the usual $SU(2)_L$ and $SU(2)_R$ indices; 
$\underline{\alpha}$ and $\underline{\dot{\alpha}}$ are indices of 
$SU(2)_{A}$ and $SU(2)_{B}$ respectively.     
\paragraph{}
To summarize, the resulting bosonic fields then have quantum numbers, 
\begin{center}
\begin{tabular}{l c c c c c}
\hline
 & $SU(2)_L$ & $SU(2)_R$ & $U(1)_{45}$ & $SU(2)_A$ & $SU(2)_B$ \\
\hline
$A_{\mu}$ & ${\bf 2}$ & ${\bf 2}$ & ${ 0}$ & ${\bf 1}$ & ${\bf 1}$ \\
$n_{\pm}$ & ${\bf 1}$ & ${\bf 1}$ & ${\pm 2}$ & ${\bf 1}$ & ${\bf 1}$ \\
$\phi_i$ & ${\bf 1}$ & ${\bf 1}$ & ${ 0}$ & ${\bf 2}$ & ${\bf 2}$ \\
\hline 
\end{tabular}
\end{center}
while the fermions transform as, 
\begin{center}
\begin{tabular}{l c c c c c}
\hline
 & $SU(2)_L$ & $SU(2)_R$ & $U(1)_{45}$ & $SU(2)_A$ & $SU(2)_B$\\
\hline
\vspace{3pt}
$\lambda^{\alpha}_{\underline{\alpha}}$ & ${\bf 2}$ & ${\bf 1}$ & ${ +1}$ & 
${\bf 2}$ & ${\bf 1}$ \\
\vspace{3pt}
$\bar{\lambda}^{\dot{\alpha}}_{\underline{\alpha}}$ & ${\bf 1}$ & ${\bf 2}$ & 
${ -1}$ & ${\bf 2}$ & ${\bf 1}$ \\
\vspace{3pt}
$\psi^{\alpha}_{\underline{\dot{\alpha}}}$ & ${\bf 2}$ & ${\bf 1}$ & ${ -1}$ & 
${\bf 1}$ & ${\bf 2}$ \\
\vspace{3pt}
$\bar{\psi}^{\dot{\alpha}}_{\underline{\dot{\alpha}}}$ & ${\bf 1}$ & ${\bf 2}$ 
& ${+1}$ & ${\bf 1}$ & ${\bf 2}$ \\ \hline
\end{tabular}
\end{center}
\paragraph{}
We now compactify the theory by replacing the $45$-plane 
by a sphere. In conventional compactification the couplings of
fields to the curvature of the sphere are determined by their 
quantum numbers under $U(1)_{45}$ which corresponds to local rotations
in the $45$-plane. In the compactification of Maldacena and 
N\'u\~nez the theory is twisted by embedding the local
rotation group into the $SU(2)_{A}\times SU(2)_{B}$ R-symmetry group
of the theory. To accomplish this we define 
Cartan subgroups $U(1)_A$ and $U(1)_{B}$ 
of $SU(2)_A$ and $SU(2)_{B}$ with corresponding generators $Q_{A}$ and
$Q_{B}$ respectively\footnote{These generators are normalised to take
  the values $Q_{A}=\pm 1$ on states in the fundamental representation
  of $SU(2)$.}. We also define the diagonal subgroup 
$U(1)_T = D(U(1)_{45} \times U(1)_A)$ with generator $Q_T = Q_{45} +
Q_A$. The vector multiplet fields then have quantum numbers,  
\begin{center}
\begin{tabular}{l c c}
\hline
  & $U(1)_A$ & $U(1)_T$ \\
\hline
$A_{\mu}$ & $0$ & ${ 0}$  \\
$n_{\pm}$ & $0$ & ${ \pm 2}$  \\
$\phi_i$ & $\pm 1$ & ${\pm 1}$ \\
$\lambda^{\alpha}_{\underline{\alpha}}$ & $\pm 1$ & $
\left( \begin{array}{c} { +2} \\ { 0} \end{array}\right)$ \\
$\bar{\lambda}^{\dot{\alpha}}_{\underline{\alpha}}$ & $\pm 1$ & $\left( 
\begin{array}{c} {0} \\ { -2} \end{array}\right)$ \\
\vspace{3pt}
$\psi^{\alpha}_{\underline{\dot{\alpha}}}$ & $0$ & ${ -1}$ \\
\vspace{3pt}
$\bar{\psi}^{\dot{\alpha}}_{\underline{\dot{\alpha}}}$ & $0$ & ${ +1}$ \\
\hline
\end{tabular}
\end{center}
\paragraph{}
We then compactify the theory with $U(1)_{T}$ playing the role of the 
local rotation group (instead of $U(1)_{45}$). We will refer to the
$U(1)_{T}$ quantum number as T-spin and the six-dimensional fields 
can be split up accordingly as,   
\begin{center}
\begin{tabular}{l l c l}
\vspace{3pt}
T-scalars: & $Q_T = 0$ && $A_{\mu}$, $\lambda^{\alpha}_{\underline{\alpha}=2}$,
$\bar{\lambda}^{\dot{\alpha}}_{\underline{\alpha}=1}$ \\
\vspace{3pt}
T-spinors: & $Q_T = \pm 1$ && $\psi^{\alpha}_{\underline{\dot{\alpha}}}$, 
$\bar{\psi}^{\dot{\alpha}}_{\underline{\dot{\alpha}}}$, $\phi_i$ \\
T-vectors: & $Q_T = \pm 2$ && $n_{\pm}$, 
$\lambda^{\alpha}_{\underline{\alpha}=1}$, 
$\bar{\lambda}^{\dot{\alpha}}_{\underline{\alpha}=2}$
\end{tabular}
\end{center}
Correspondingly the terms scalar, spinor and vector 
will be reserved for describing
the transformation properties of fields under the four-dimensional 
Lorentz group. The existence of a single Weyl spinor (of both
chiralities) which is also a T-scalar guarentees the existence of a 
single massless fermion in four-dimensions as required by ${\cal N}=1$
supersymmetry. 

\subsection{Maldacena-N\'u\~nez Bosonic Action}\label{sec:MNA}

A direct calculation of the MN action can be performed by the twisted 
compactification of the $\mathcal{N}=(1,1)$ SUSY Yang-Mills action, allowing 
a comparison with the effective six-dimensional action of the Higgsed 
$\mathcal{N}=1^*$ theory. In this section we will calculate the bosonic part 
of this action. The $\mathcal{N}=(1,1)$ action with gauge group $U(1)$ is 
obtained via the trivial dimensional reduction of the $\mathcal{N}=1$ SUSY 
Yang-Mills theory in ten spacetime dimensions with gauge group $U(1)$ (see 
Appendix \ref{app:1} for more details). The action for this $U(1)$ theory is,
\begin{equation}
\mathcal{S} = \frac{1}{g^2_6} \int d^6 x \left\lbrace -\frac{1}{4} \, F_{ij} 
F^{ij} - \frac{1}{2} \, \partial_i \phi_n \partial^i \phi^n \right\rbrace
\end{equation}
where $i,j = 0,1, \dots, 5$ and $n = 1,\dots, 4$. We split the manifold 
$\Re^{5,1} \to \Re^{3,1} \times \Re^2$, and let $\mu = 0,1,2,3$ and $a=1,2$.
\begin{equation*}
\begin{split}
\mathcal{S} & = \frac{1}{g^2_6} \int d^6 x \left\lbrace -\frac{1}{4} 
F_{\mu \nu} F^{\mu \nu} - \frac{1}{2} F_{\mu a} F^{\mu a} - \frac{1}{4} F_{ab}
F^{ab} - \frac{1}{2} \partial_{\mu} \phi_m \partial^{\mu} \phi^m \right. \\
& \qquad \left. - \frac{1}{2} \partial_a \phi_m \partial^a \phi^m 
\right\rbrace \\
& = \frac{1}{g^2_6} \int d^6 x \left\lbrace -\frac{1}{4} F_{\mu \nu} 
F^{\mu \nu} - \frac{1}{2} \partial_{\mu} n_a \partial^{\mu} n^a - 
\partial_{\mu} n_a \partial^a A^{\mu} - \frac{1}{2} \partial_a A_{\mu} 
\partial^a A^{\mu} \right. \\
& \qquad \left. - \frac{1}{4} F_{ab} F^{ab}  - \frac{1}{2} \partial_{\mu} 
\phi_m \partial^{\mu} \phi^m - \frac{1}{2} \partial_a \phi_m \partial^a \phi^m
\right\rbrace
\end{split}
\end{equation*}
We want to perform the twisted compactification on the sphere $S^2$. $U(1)_T 
\sim SO(2)_T$ is the local rotation group in the twisted compactification. From
the group structure in section 3 we see that $A_{\mu}$ is a 
T-scalar, $\phi_m$ form T-spinors and $n_a$ form T-vectors. In moving from a 
flat spacetime to a curved spacetime, derivatives on the flat spacetime become 
general covariant derivatives. $\partial_{\mu} \to \partial_{\mu}$ and 
$\partial_a \to \nabla_a$. 
\begin{subeqnarray}
F_{\mu \nu}  & \to & F_{\mu \nu} \\
\partial_{\mu} n_a & \to & \partial_{\mu} n_a 
\end{subeqnarray}
A general covariant derivative's action on a scalar is that of an ordinary 
derivative.
\begin{equation}
\partial_a A_{\mu} \ \to \ \nabla_a A_{\mu} = \partial_a A_{\mu} 
\end{equation}
For a vector $\partial_a n_b \to \partial_a n_b - \Gamma^c_{ab} n_c$, where 
$\Gamma^c_{ab}$ is a Christoffel symbol.
\begin{equation}
\begin{split}
F_{ab} = \partial_a n_b - \partial_b n_a \ \to \ \mathcal{F}_{ab} & 
= \partial_a n_b - \Gamma^c_{ab} n_c - \partial_b n_a + \Gamma^c_{ba} n_c \\
& = \partial_a n_b - \partial_b n_a 
\end{split}
\end{equation}

Consider the real scalars $\phi_m$ of $SO(4)$. In order to transform these real
scalars into T-spinors on the sphere we need to reveal the spin structure of 
$SU(2)_A$. We construct an $SO(4)$ bispinor,
\begin{eqnarray*}
v_{\underline{\alpha} \underline{\dot{\alpha}}} & = & 
i(\tau^m)_{\underline{\alpha} \underline{\dot{\alpha}}} \phi_m \\
\phi^m & = & -\frac{i}{2} (\bar{\tau}^m)^{\underline{\dot{\alpha}} 
\underline{\alpha}} v_{\underline{\alpha} \underline{\dot{\alpha}}}
\end{eqnarray*}
Substituting this expression for the real scalars into the last term of the 
action,
\begin{eqnarray*}
\int d^2 x \, \partial_a \phi_m \partial^a \phi^m & = & -\frac{1}{4} \int d^2 x
 \, \partial_a \left( (\bar{\tau}^m)^{\underline{\dot{\alpha}} 
 \underline{\alpha}} v_{\underline{\alpha} \underline{\dot{\alpha}}} \right) 
 \partial^a \left( (\bar{\tau}_m)^{\underline{\dot{\beta}} \underline{\beta}} v_{\underline{\beta} \underline{\dot{\beta}}} \right) \\
& = & \frac{1}{2} \int d^2 x \, \partial_a v^{\underline{\alpha} 
\underline{\dot{\alpha}}} \partial^a v_{\underline{\alpha} 
\underline{\dot{\alpha}}} \\
& = & \frac{1}{2} \int d^2 x \, v^{\underline{\alpha}}_{\phantom{\alpha} \underline{\dot{\alpha}}} \partial_a \partial^a v^{\phantom{\alpha} \underline{\dot{\alpha}}}_{\underline{\alpha}} \\
& = & -\frac{1}{2} \int d^2 x \, \Xi^{\dag}_{\hat{\alpha}} (\partial_a 
\partial^a \otimes \mathbb{1}_2)^{\hat{\alpha}}_{\phantom{\alpha} \hat{\beta}}
\, \Xi^{\hat{\beta}}
\end{eqnarray*}
where $\Xi^{\hat{\alpha}} = \left( \begin{array}{c} v^{\phantom{\alpha} \underline{\dot{\alpha}}}_{\underline{1}} \\ v^{\phantom{\alpha} \underline{\dot{\alpha}}}_{\underline{2}} \end{array} \right)$ and $(v_{\underline{\alpha}}^{\phantom{\alpha} \underline{\dot{\alpha}}})^{\dag} 
= (\lambda^A_{\underline{\alpha}} \bar{\lambda}^{\underline{\dot{\alpha}}}_A)
^{\dag} = - v^{\underline{\alpha}}_{\phantom{\alpha} \underline{\dot{\alpha}}}$
. The differential operator can be rewritten,
\begin{equation}
\partial_a \partial^a \otimes \mathbb{1}_2 = \sigma^a \partial_a \sigma^b 
\partial_b = \partial\!\!\!/^2
\end{equation}
Moving from flat spacetime to curved spacetime, $\partial\!\!\!/ \to 
\hat{\nabla}_{S^2}$,
\begin{equation}
-\frac{1}{2} \, \Xi^{\dag}_{\hat{\alpha}} 
(\partial\!\!\!/^2)^{\hat{\alpha}}_{\phantom{\alpha} \hat{\beta}} 
\Xi^{\hat{\beta}} \ \to \ -\frac{1}{2} \, \Xi^{\dag}_{\hat{\alpha}} (\hat{\nabla}^2_{S^2})^{\hat{\alpha}}_{\phantom{\alpha} \hat{\beta}} \, 
\Xi^{\hat{\beta}} = \frac{1}{2} \Xi^{\dag}_{\hat{\alpha}} [(-i \hat{\nabla}_{S^2})^2]^{\hat{\alpha}}_{\phantom{\alpha} \hat{\beta}} \, 
\Xi^{\hat{\beta}}
\end{equation}
Similarly,
\begin{equation}
\partial_{\mu} \phi_m \partial^{\mu} \phi^m = \frac{1}{2} \partial_{\mu} \Xi^{\dag}_{\hat{\alpha}} \partial^{\mu} \Xi^{\hat{\alpha}}
\end{equation}
Therefore the bosonic MN action is,
\begin{equation}\label{result}
\begin{split}
\mathcal{S}_B = \frac{1}{g^2_6} & \int d^4 x \, \int R^2 d\Omega \, \left\lbrace 
-\frac{1}{4} \, F_{\mu \nu} F^{\mu \nu} - \frac{1}{2} \, \partial_{\mu} n_a 
\partial^{\mu} n^a - \partial_{\mu} n_a \partial^a A^{\mu}  \right. \\
& \left. - \frac{1}{2} \,
\partial_a A_{\mu} \partial^a A^{\mu} - \frac{1}{4} \, \mathcal{F}_{ab} \mathcal{F}^{ab}
- \frac{1}{4} \, \partial_{\mu} \Xi^{\dag}_{\hat{\alpha}} \partial^{\mu} \Xi^{\hat{\alpha}}
- \frac{1}{4} \, \Xi^{\dag}_{\hat{\alpha}} [(-i \hat{\nabla}_{S^2})^2]^{\hat{\alpha}}
_{\phantom{\alpha} \hat{\beta}} \, \Xi^{\hat{\beta}} \right\rbrace
\end{split}
\end{equation}

\subsection{Maldacena-N\'u\~nez Kaluza-Klein Spectrum}\label{sec:MNs}

Each six-dimensional field has a kinetic term on
$S^{2}$. 
After expanding in appropriate spherical harmonics, this
kinetic term determines the masses of an infinite tower of
four-dimensional fields. 
We now consider the Kaluza-Klein spectrum of each type
of field in turn.  
\paragraph{}
After integration by parts, the kinetic term for a T-scalar field 
$A_{\mu}$ defined on a 2-sphere of unit radius can be written as,  
\begin{equation}
\mathcal{S}_{A}= \frac{1}{g^2_6} \, \int \, R^2 d \Omega \,\, A_{\mu} 
\Delta_{S^{2}} A^{\mu} 
\end{equation}
To find the mass eigenstates we expand $A_{\mu}$ 
in terms of spherical harmonics as, 
\begin{eqnarray}
 A_{\mu}(\theta,\phi) & = & 
\sum^{\infty}_{l=0} \sum^{+l}_{m=-l} A_{(\mu) lm} Y_{lm}(\theta,\phi) \\
\nonumber 
\end{eqnarray}
From section \ref{sec:sphere} we know the eigenvalues of the scalar
Laplacian are $\mu \sim l(l+1)$ for integer $l\geq 0$ and for 
$m=-l,\ldots +l$. Thus for each T-scalar field in six dimensions we
find a Kaluza-Klein tower of four-dimensional 
fields with masses,
\begin{equation}
M^2 = \frac{1}{R^2} \, l(l+1)
\end{equation} 
with degeneracy $(2l+1)$. According to the 
list of T-scalar fields given above we find a four-dimensional vector
field, a left-handed Weyl spinor and a right-handed Weyl spinor at each 
mass level. The corresponding representations of the four-dimensional 
Lorentz group $SU(2)_{L}\times SU(2)_{R}$ are,
\begin{eqnarray*}
(\mathbf{2},\mathbf{2}) \oplus (\mathbf{2},\mathbf{1}) 
\oplus (\mathbf{1},\mathbf{2})
\end{eqnarray*} 
For $l=0$, the corresponding four-dimensional fields are massless 
and the spin quantum numbers match those of 
a single massless vector multiplet of ${\cal N}=1$ supersymmetry.   
For $l>1$, we find massive vector fields together with Weyl 
fermions. However, a massive vector multiplet of ${\cal N}=1$
supersymmetry also includes scalar fields in four-dimensions. 
Thus, for $l>0$, the fields descending from the T-scalars in six dimensions 
do not form complete multiplets of ${\cal N}=1$ supersymmetry. 
This puzzle will be resolved below where we will find the additional 
states needed to form massive vector multiplets.    
\paragraph{}
A two-component Dirac T-spinor $\Upsilon$ defined on a 2-sphere has
kinetic term, 
\begin{equation}
\mathcal{S}_{\Upsilon} = \frac{1}{g^2_6} \, \int\, R^2 d \Omega\, \, i 
\bar{\Upsilon} \, \hat{\nabla}_{S^2} \Upsilon 
\label{dirac2}
\end{equation}
From section \ref{sec:sphere} we know the eigenvalues of the Dirac operator on 
the 2-sphere correspond to $\mu \sim l^{2}$ for 
integer $l\geq 1$ with degeneracy $2l$. 
The fermionic T-spinor fields, $\psi^{\alpha}_{\underline{\dot{\alpha}}}$ and 
$\bar{\psi}^{\dot{\alpha}}_{\underline{\dot{\alpha}}}$ listed above can be 
combined to form two-component Dirac spinors on $S^{2}$ with kinetic
terms (\ref{dirac2}) according to, 
\begin{equation}
\Upsilon^{\alpha}_{(\underline{\dot{\alpha}})} = \left(\begin{array}{l} 
\vspace{3pt} 
\psi^{\alpha}_{\underline{\dot{\alpha}}} \\
\bar{\psi}^{\dot{\alpha}=\alpha}_{\underline{\dot{\alpha}}}
\end{array}\right)
\end{equation}
for $\underline{\dot{\alpha}}=1,2$, $\alpha=1,2$. Thus we obtain four-species 
of Dirac
spinors on the 2-sphere. Each species yields $2l$ states of mass,
\begin{equation} 
M^2 = \frac{1}{R^2} \, l^{2}
\end{equation}
for $l\geq 1$ after expansion in terms of eigenstates of
the squared Dirac operator. At each mass level we therefore find
$8l$ off-shell degrees of freedom which can be recombined as $4l$
left-handed and $4l$ right-handed Weyl spinors in four dimensions. 
These Weyl spinors must 
be paired with bosonic fields to form multiplets of 
${\cal N}=1$ SUSY in four dimensions. 
The extra fields come from Kaluza-Klein reduction of the bosonic T-spinors 
$\phi_{i}$, $i=1,2,3,4$, which yield massive scalar fields in four
dimensions. These states combine with the fermionic T-spinors to form 
massive chiral multiplets with Lorentz spins,  
\begin{eqnarray*}
(\mathbf{2},\mathbf{1}) \oplus (\mathbf{1},\mathbf{2}) \oplus 
2 \times (\mathbf{1},\mathbf{1})
\end{eqnarray*}
\paragraph{}
It remains to determine the Kaluza-Klein spectrum of the T-vector
fields. As we have unbroken ${\cal N}=1$ supersymmetry in the
four non-compact dimensions it suffices to focus on the 
bosonic T-vector fields $n_{\pm}=(A_{4}\pm iA_{5})/\sqrt{2}$. 
The two real components $A_{4}$ and $A_{5}$ define a Maxwell gauge 
field $n_{a}$ on the 2-sphere. The resulting 
kinetic term reads, 
\begin{equation}
\mathcal{S}_{n} = \frac{1}{g^2_6} \, \int R^2 d \Omega\,\,\frac{1}{4} \mathcal{F}_{ab} 
\mathcal{F}^{ab}
\end{equation}
where $\mathcal{F}_{ab}=
\partial_{a}n_{b}-\partial_{b}n_{a}$. 
We impose the gauge condition $\nabla^a n_a = 0$ and expand
${\bf n}=(n_{\theta},n_{\phi})$ in terms of vector spherical harmonics,   
\begin{eqnarray}
\nonumber {\bf n} & = & \sum_{l,m} n_{lm} {\bf T_{lm}} 
\end{eqnarray}
The field tensor expands as,
\begin{equation}
\begin{split}
\mathcal{F}_{\theta \phi} & = R \, \sum_{l,m} n_{lm} 
\frac{1}{\sqrt{l(l+1)}} \Big( \partial_{\theta} (\textrm{sin} \, \theta \, 
\partial_{\theta} Y_{lm}) + \textrm{csc} \, \theta \, \partial_{\phi} 
\partial_{\phi} Y_{lm} \Big) \\
& = R \, \sum_{l,m} n_{lm} \frac{1}{\sqrt{l(l+1)}} \, \textrm{sin} \, \theta \, 
L^2 Y_{lm}
\end{split}
\end{equation}
The Maxwell term is,  
\begin{eqnarray}
\mathcal{S}_{n} = \frac{1}{g^2_6} \, \sum_{l,m,l',m'} 
n^{\dag}_{lm} n^{\phantom{\dag}}_{l'm'} \, l(l+1) \, \delta_{ll'} \delta_{mm'}
\end{eqnarray}
Thus the T-vector field $n_{\pm}$ yields a Kaluza-Klein tower of 
four-dimensional scalar fields of mass,
\begin{equation}
M^2 = \frac{1}{R^2} \, l(l+1)
\end{equation}
with degeneracy $2l+1$
for $l\geq 1$. Notice that these fields are degenerate in mass with
the four-dimensional vector fields coming from the Kaluza-Klein
reduction of the T-scalars. In fact the number of scalar fields is
just right to pair up with the massive vector fields to form massive 
vector multiplets of ${\cal N}=1$ SUSY in four dimensions with Lorentz
spins,  
\begin{eqnarray*}
(\mathbf{2},\mathbf{2}) \oplus 2 \times [(\mathbf{2},\mathbf{1}) 
\oplus (\mathbf{1},\mathbf{2})] \oplus (\mathbf{1},\mathbf{1})
\end{eqnarray*}
The fermionic part of each multiplet includes two species of left- and
right-handed Weyl fermions. One species comes from the KK reduction
of the fermionic T-scalars  $\lambda^{\alpha}_{\underline{\alpha}=2}$ and
$\bar{\lambda}^{\dot{\alpha}}_{\underline{\alpha}=1}$ and the other comes from
the reduction of the fermionic T-vectors 
$\lambda^{\alpha}_{\underline{\alpha}=1}$ and
$\bar{\lambda}^{\dot{\alpha}}_{\underline{\alpha}=2}$. 
\paragraph{}
We summarise the complete Kaluza-Klein spectrum of the MN
compactification in the table below, 
\begin{center}
\end{center}
T-scalar:
\begin{center}
\begin{tabular}{c|c}
$\lambda$ & States \\
\hline
$l(l+1)$ & $(2l+1) \times$ $\Big\lbrace ({\bf 2},{\bf 2}) \oplus 
({\bf 2},{\bf 1}) \oplus ({\bf 1}, {\bf2}) \Big\rbrace$
\end{tabular}\\
$l = 0,1,2,\dots$ \qquad
\end{center}
T-spinor:
\begin{center}
\begin{tabular}{c|c}
$\lambda$ & States \\
\hline
$l^2$ & $4l \times$ $\Big\lbrace ({\bf 2},{\bf 1}) \oplus ({\bf 1},{\bf 2}) 
\oplus 2 \times ({\bf 1},{\bf 1}) \Big\rbrace$
\end{tabular}\\
$l = 1,2,3,\dots$
\end{center}
T-vector:
\begin{center}
\begin{tabular}{c|c}
$\lambda$ & States \\
\hline
$l(l+1)$ & $(2l+1) \times$ $\Big\lbrace ({\bf 2},{\bf 1}) \oplus 
({\bf 1},{\bf 2}) \oplus 2 \times ({\bf 1},{\bf 1}) \Big\rbrace$
\end{tabular}\\
$l = 1,2,3,\dots$
\end{center}
\paragraph{}
We can also present the spectrum in terms of complete ${\cal N}=1$
multiplets. The spectrum includes a single massless $U(1)$ vector
multiplet. The massive spectrum is labelled by a positive integer
$l=1,2,\ldots$ and includes the following states,  
\begin{center}
\begin{tabular}{c c l}
\hline
Mass & Degeneracy & Multiplet \\
\hline
\vspace{3pt}
$\frac{1}{R^2} \, l(l+1)$ & $2l+1$ & Massive Vector \\
\vspace{3pt}
$\frac{1}{R^2} \, l^{2}$   & $4l$  & Massive Chiral \\
\hline
\vspace{3pt}
\end{tabular}
\end{center}

\section{Deconstruction}\label{sec:decon}

In this Section we motivate the appearance of extra dimensions in the 
large-$N$ limit of $\mathcal{N}=1^*$ SUSY Yang-Mills with $U(N)$ gauge
group. The 
$\mathcal{N}=1^*$ theory is a relevant deformation of the $\mathcal{N}=4$ SUSY
Yang-Mills theory. We deform the $\mathcal{N}=4$ theory's superpotential by 
adding mass terms for the chiral multiplets,
\begin{equation}\label{eq:superpotential}
\mathcal{W}(\Phi) = \textrm{Tr}_N \left( i\sqrt{2} \, \Phi_1 [\Phi_2, \Phi_3] +
\eta \sum_{i=1}^{3}\Phi^2_i \right)
\end{equation}
The theory has no moduli space, instead it contains a number of isolated vacua 
\cite{VW}. The F-flatness condition is,
\begin{equation}\label{eq:vacua}
[\Phi_i, \Phi_j] = \sqrt{2} \eta \, i \varepsilon_{ijk} \Phi_k
\end{equation}
Under the reparameterization,
\begin{equation}\label{eq:repar}
\Phi_i \to \frac {1}{\sqrt{2} \eta} \Phi_i
\end{equation}
the F-flatness condition (\ref{eq:vacua}) becomes,
\begin{equation}\label{eq:vacua1}
[\Phi_i, \Phi_j] = i\varepsilon_{ijk} \Phi_k
\end{equation}
which is precisely the $SU(2)$ Lie algebra. It can be solved by any 
$N$-dimensional representation of the $SU(2)$ generators, which in general will
be reducible. This solution also satisfies the D-flatness condition. Our gauge
group is $U(N) \sim SU(N) \times U(1)$, with $\Phi_i$ represented by 
$N \times N$ matrices.
There is a single irreducible representation $J^{(d)}_i$ of the $SU(2)$ Lie 
algebra for every dimension $d$, which allows the gauge group to be decomposed 
into a number of irreducible representations, of total dimension $N$.
If the number of times a representation $d$ appears is denoted $k_d$, then the 
unbroken gauge group is $U(N) \to \otimes_d U(k_d)$. We are interested in the 
Higgs vacuum where $\Phi_i = J^{(N)}_i$, breaking the gauge group 
$U(N) \to U(1)$. More general Higgs branches are present, where the gauge group
is broken $U(N=pq) \to U(p)$ by $p$ copies of the $q$-dimensional 
representation of the $SU(2)$ Lie algebra, 
$\Phi_i = \mathbb{1}_p \otimes J^{(q)}_i$.
\paragraph{}
The extra dimensions emerge via the mechanism seen in M(atrix) theory, which we
will illustrate in the rest of this section. It was interpreted as 
deconstruction in \cite{AF}. We find that the Higgs vacuum describes a fuzzy 
sphere \cite{bib:Madore}. If we rescale the matrix fields\footnote{We denote 
all matrix fields with hats.} $\hat{x}_i = \tau \hat{\Phi}_i$, where $\tau^2 =
 \frac{4R^2}{N^2 - 1}$, then the expectation values in the Higgs vacuum satisfy,
\begin{equation}
\hat{x}_{1}^{2}+\hat{x}_{2}^{2}+ \hat{x}_{3}^{2}= \mathbb{1}
\label{ncs2}
\end{equation}
which is the defining equation of a fuzzy sphere. The coordinates of the fuzzy 
sphere $\hat{x}_i$ have the commutation relation,
\begin{equation}
[\hat{x}_i, \hat{x}_j] = i \tau \, \varepsilon_{ijk} \hat{x}_k
\end{equation}
We recover the ordinary commutative sphere in the limit $N \to \infty$.

The Higgsed $\mathcal{N}=1^*$ theory is a theory of $N\times N$
matrices. There is a well known correspondence between such 
matrix theories and 
non-commutative field theories. Scalar functions on a 2-sphere can be expanded 
in terms of spherical harmonics,
\begin{equation}\tag{\ref{eq:expand}}
a(\theta, \phi) = \sum^{\infty}_{l=0} \sum^l_{m=-l} a_{lm}
Y_{lm}(\theta, \phi)
\end{equation}
The spherical harmonics can be expressed in terms of the cartesian
coordinates $x_{A}$ with $A=1,2,3$ of a unit vector in $\Re^{3}$ 
\cite{bib:Kimura},  
\begin{equation}\label{eq:spharm}
Y_{lm}(\theta, \phi) = \sum_{\vec{A}} 
f ^{(lm)} _{A_1 \dots A_l} x^{A_1} \! \dots x^{A_l}
\end{equation}
where $f ^{(lm)} _{A_1 \dots A_l}$ is a traceless symmetric tensor of 
$SO(3)$ with rank $l$.
Similarly we can expand $N \times N$ matrices of a matrix theory on a fuzzy 
sphere as,
\begin{eqnarray}\label{eq:fuzzyexpand}
\hat{a} & = & \sum^{N-1}_{l=0} \sum^l_{m=-l} a_{lm} \hat{Y}_{lm} \\
\hat{Y}_{lm} &
 = & R^{-l} \sum_{\vec{A}} f ^{(lm)} _{A_1 \dots A_l} 
\hat{x}^{A_1} \! \dots \hat{x}^{A_l}
\end{eqnarray}
where $\hat{x}_{A}=\frac{2R}{\sqrt{N^{2}-1}} \, J^{(N)}_{A}$ and 
$f ^{(lm)}_{A_1 \dots A_l}$ is the same tensor as in \eqref{eq:spharm}. 
The matrices $\hat{Y}_{lm}$ are known as 
fuzzy spherical harmonics. They obey the orthonormality condition,  
\begin{equation}
\textrm{Tr}_N \left( \hat{Y}^{\dag}_{lm} \hat{Y}_{l'm'}^{\phantom{\dag}} 
\right) = 
\delta_{l l'} \, \delta_{m m'}
\end{equation}
There is an obvious relation between equations (\ref{eq:expand}) and 
(\ref{eq:fuzzyexpand}).
\begin{equation}\label{eq:map}
\hat{a} = \sum^{N-1}_{l=0} \sum^{l}_{m=-l} a_{lm} \hat{Y}_{lm} \to
\ a(\theta, \phi) = 
\sum^{N-1}_{l = 0} \sum^l_{m = -l} a_{lm} Y_{lm}(\theta, \phi)
\end{equation}
Notice that the expansion in spherical harmonics is truncated at 
$N-1$ reflecting the finite number of degrees of freedom in the matrix 
$\hat{a}$. This is a 1:1 mapping, formally given by \cite{bib:Kimura},
\begin{equation}
a(\theta, \phi) = \sum_{lm} \textrm{Tr}_N (\hat{Y}^{\dag}_{lm} \hat{a}) Y_{lm}
(\theta, \phi)
\end{equation}
The matrix trace is mapped by equation \eqref{eq:map} to an integral over the 
sphere.
\begin{equation}\label{eq:maptr}
\frac{1}{N} \textrm{Tr}_N \to \frac{1}{4 \pi} \int d\Omega
\end{equation}
The product of matrices maps to the star-product on the non-commutative sphere 
\footnote{In order for the mapping to remain 1:1 we must assume that N is 
sufficiently large such that $l + l' \ngtr N-1$.},
\begin{equation}
a \ast b(\theta, \phi) = \sum_{lm} \textrm{Tr}_N (\hat{Y}^{\dag}_{lm} \hat{a} 
\hat{b}) Y^{\phantom{\dag}}_{lm}(\theta, \phi)
\end{equation}
This product is non-commutative due to the non-commutative nature of
matrix multiplication. 
This mapping produces a correspondence between matrix theories and 
non-commutative field theories.
\paragraph{}
In analogy to continuum field theory we have derivative operators for the 
matrix theory. They correspond to the adjoint action of $J^{(N)}_i$ 
\cite{bib:Kimura}.
\begin{subeqnarray}\label{eq:adaction}
Ad(J^{(N)}_3) \ & = &  \ \sum_{lm} a_{lm} \left[ J^{(N)}_3, \hat{Y}_{lm} 
\right] \,  = \ \sum_{lm} a_{lm} \, m \, \hat{Y}_{lm} \\
Ad(J^{(N)}_{\pm}) \ & = &  \ \sum_{lm} a_{lm} \left[ J^{(N)}_{\pm}, 
\hat{Y}_{lm} \right]  \\
\nonumber & = & \ \sum_{lm} a_{lm} \sqrt{(l \pm m +1)(l \mp m)} \, 
\hat{Y}_{lm \pm 1} \;
\end{subeqnarray}
The properties above, equations \eqref{eq:adaction}, show that by the 
correspondence between matrices and functions \eqref{eq:map} the adjoint action
 of $J^{(N)}_i$ becomes,
\begin{equation}
Ad(J^{(N)}_i) \to L_i 
\end{equation}
The operator $L_i$ is the derivative operator on the non-commutative sphere.

The fuzzy spherical harmonics have the commutator \cite{bib:hoppe},
\begin{equation}\label{eq:harcom}
\left[ Y_{l_1 m_1}, Y_{l_2 m_2} \right] = F^{l_3 m_3}_{l_1 m_1 l_2 m_2} 
Y_{l_3 m_3}
\end{equation}
where the structure constants are,
\begin{equation}
\begin{split}
F^{l_3 m_3}_{l_1 m_1 l_2 m_2} = 2 & \sqrt{(2l_1+1)(2l_2+1)(2l_3+1)}(-1)^{N-1} \\
& \times \left(\begin{array}{ccc} l_1 & l_2 & l_3 \\ m_1 & 
m_2 & m_3 \end{array}\right) \left\lbrace\begin{array}{ccc} l_1 & l_2 & l_3 \\ 
\frac{N-1}{2} & \frac{N-1}{2} & \frac{N-1}{2} \end{array} \right\rbrace
\end{split}
\end{equation}
$(\dots)$ is a Wigner 3j-symbol and $\lbrace \dots \rbrace$ is a Wigner 
6j-symbol.  For large-$N$ the 6j-symbol behaves as $N^{-3/2}$ \cite{bib:hoppe}.
In the limit $N \to \infty$ the commutator \eqref{eq:harcom} becomes,
\begin{equation}\label{eq:nocom}
\left[ Y_{l_1 m_1}, Y_{l_2 m_2} \right] = 0
\end{equation}
and we recover the usual commutative spherical harmonics, this is the 
commutative limit.

The correspondence \eqref{eq:map} allows us to map a matrix model to a 
non-commutative field theory.
For the Higgsed $\mathcal{N}=1^*$ theory this mechanism, from M(atrix) theory 
\cite{bib:Kimura, bib:WATI}, produces a six-dimensional non-commutative field 
theory, with a UV cutoff for finite N. The limit $N \to \infty$ the theory 
becomes a commutative, continuum field theory on $\Re^{3,1} \times S^2$.

\section{Classical $\mathcal{N}=1^*$ SUSY Yang-Mills Spectrum}\label{sec:mass}

In this section we will calculate the full classical spectrum of the Higgsed 
$\mathcal{N}=1^*$ theory. The classical action of $\mathcal{N}=1^*$ theory 
corresponding to the superpotential (\ref{eq:superpotential}) with 
reparameterization (\ref{eq:repar}) is,
\begin{equation}
\begin{split}
\mathcal{S} = & \frac{1}{g^2_{ym}} \int d^4 x \, \textrm{Tr}_N 
\bigg\lbrace -\frac{1}{4} \, \eta^2 \, F_{\mu \nu} F^{\mu \nu} - i \eta^3 \, 
\lambda \sigma^{\mu} D_{\mu} \bar{\lambda} - i \eta^3 \, \psi_i 
\sigma^{\mu} D_{\mu} \bar{\psi}_i  \\
& -2 \eta^2 D_{\mu} \Phi^\dag _i D^{\mu} \Phi_i + \eta^4 \Big( i \psi_i 
[\Phi^{\dag}_i , \lambda ] - i \lambda [ \Phi^{\dag}_i , \psi_i ] 
+ i \bar{\psi}_i [ \Phi_i , \bar{\lambda} ] - i \bar{\lambda} 
[ \Phi_i , \bar{\psi}_i ] \\
& + i \psi_i \, \varepsilon_{ijk} [ \Phi_k, \psi_j ] + 
i \bar{\psi}_i \, \varepsilon_{ijk} [ \Phi^{\dag}_k, \bar{\psi}_j  ] - 
\psi_i \psi_i - \bar{\psi}_i \bar{\psi}_i
- 2[\Phi_i , \Phi^{\dag}_i ]^2 \\
& + 4 [\Phi^{\dag}_i , \Phi^{\dag}_j ][\Phi_i , \Phi_j ] - 4i 
\varepsilon_{ijk} \Phi^{\dag}_i [\Phi_j , \Phi_k ] - 4i \varepsilon_{ijk} 
[\Phi^{\dag}_i , \Phi^{\dag}_j ]\Phi_k - 8 \, \Phi^{\dag}_i \Phi_i \Big) 
\bigg\rbrace
\end{split}
\end{equation}
where $\lambda$ is the gaugino and $\psi_i$ are the superpartners of $\Phi_i$; 
$F_{\mu \nu} = \partial_{\mu} A_{\nu} - \partial_{\nu} A_{\mu} + i\eta 
[A_{\mu}, A_{\nu}]$ and $D_{\mu} \Phi = \partial_{\mu} \Phi + 
i\eta[A_{\mu}, \Phi]$. 
We have chosen to normalise all our fields such that every term in the action 
has the prefactor $1/g^2_{ym}$. We expand the complex scalars about the Higgs 
vacuum $\hat{\Phi}_i = J^{(N)}_i + \delta\hat{\Phi}_i$, only terms quadratic 
in the fields contribute to the mass spectrum, hence we will ignore the higher 
orders. We will derive the spectrum from both the fermionic and bosonic 
contributions. We first calculate the fermionic mass matrix, the contribution 
is,
\begin{equation}\label{eq:ferintmass}
\begin{split}
\mathcal{L}_{FM} = 2 \eta \, \textrm{Tr}_N \Big\lbrace & i\hat{\psi}_i \, 
\varepsilon_{ijk}[J_k , \hat{\psi}_j] + i \hat{\bar{\psi}}_i \varepsilon_{ijk} 
\, [J_k , \hat{\bar{\psi}}_j] - i\hat{\lambda} [ J_i , \hat{\psi}_i ] \\ 
& + i\hat{\psi}_i [J_i , \hat{\lambda} ] + i\hat{\bar{\psi}}_i 
[ J_i , \hat{\bar{\lambda}} ] - i\hat{\bar{\lambda}} [ J_i , 
\hat{\bar{\psi}}_i ] - \hat{\psi}_i \hat{\psi}_i - \hat{\bar{\psi}}_i 
\hat{\bar{\psi}}_i
\Big\rbrace
\end{split}
\end{equation}
where we have suppressed the label $(N)$. We rewrite this expression to form 
the fermionic mass matrices,
\begin{equation}
\mathcal{L}_{FM} = 2 \eta \Big\lbrace \big(\hat{\Psi}_R \big)_{ab} 
\Delta^{(RS)}_{ab,cd} \big(\hat{\Psi}^T_S \big)_{cd} + 
\big(\hat{\bar{\Psi}}_R \big)_{ab} \bar{\Delta}^{(RS)}_{ab,cd} 
\big(\hat{\bar{\Psi}}^T_S \big)_{cd} \Big\rbrace
\end{equation}
where the four species of Weyl fermion are combined in a column
$\hat{\Psi}_{R}$ with $\hat{\Psi}_R = \hat{\psi}_i$ for $R=i=1,2,3$ and 
$\hat{\Psi}_4 =
\hat{\lambda}$. The mass matrices $\Delta$ and $\bar{\Delta}$ have the explicit
form,
\begin{subeqnarray}
\Delta^{(ij)}_{ab,cd} & = & i\varepsilon_{ijk} \Big(\delta_{ac} (J_k)_{bd} - 
(J^*_k)_{ac} \, \delta_{bd} \Big) - \delta_{ij} \, \delta_{ac} \, \delta_{bd} 
= \bar{\Delta}^{(ij)}_{ab,cd} \\
\Delta^{(i4)}_{ab,cd} & = & -i\Big((J^*_i)_{ac} \, \delta_{bd} - \delta_{ac} 
(J_i)_{bd}\Big) = \bar{\Delta}^{(i4)}_{ab,cd} \\
\Delta^{(4i)}_{ab,cd} & = & -i\Big(\delta_{ac} (J_i)_{bd} - (J^*_i)_{ac} \, 
\delta_{bd}\Big) = \bar{\Delta}^{(4i)}_{ab,cd}
\end{subeqnarray}
The masses of physical states are detemined by the squared mass matrix, 
\begin{equation}
M^{(R S)}_{ab, \, ef} = 4 \eta^2
\bar{\Delta}^{(R T)}_{ab, \, cd} \, \Delta^{(T S)}_{cd, \, ef}
\end{equation}
Explicitly these matrices are,
\begin{subeqnarray}
\nonumber M^{(ij)}_{ab,ef} & = & (J^*_i J^*_j - J^*_j J^*_i)_{ae} \delta_{bf} +
 \delta_{ae} (J_i J_j - J_j J_i)_{bf} \\
& & - 2i\varepsilon_{ijk} \Big( \delta_{ae}(J_k)_{bf} - (J^*_k)_{ae} 
\delta_{bf} \Big) + \delta_{ij} \Big\lbrace \delta_{ae}(J_k J_k)_{bf} \\
\nonumber & & - 2(J^*_k)_{ae} (J_k)_{bf} + (J^*_k J^*_k)_{ae} \delta_{bf} + 
\delta_{ae} \delta_{bf} \Big\rbrace \\
M^{(i4)}_{ab,ef} & = & \varepsilon_{ijk} \Big( (J^*_j)_{ae} (J_k)_{bf} - 
(J^*_k J^*_j)_{ae} \delta_{bf} 
- \delta_{ae} (J_k J_j)_{bf} \\
\nonumber & & + (J^*_k)_{ae}(J_j)_{bf} \Big)  + i \Big( (J^*_i)_{ae} 
\delta_{bf} - \delta_{ae} (J_i)_{bf} \Big) \\
M^{(4i)}_{ab,ef} & = & -\varepsilon_{ijk} \Big\lbrace \delta_{ae}(J_j J_k)_{bf}
 - (J^*_k)_{ae} (J_j)_{bf}
- (J^*_j)_{ae} (J_k)_{bf} \\
\nonumber & & + (J^*_j J^*_k)_{ae} \delta_{bf} \Big\rbrace  - i\Big( 
(J^*_i)_{ae} \delta_{bf} - \delta_{ae} (J_i)_{bf} \Big) \\
M^{(44)}_{ab,ef} & = & \delta_{ae} (J_i J_i)_{bf} - 2(J^*_i)_{ae} (J_i)_{bf} 
+ (J^*_i J^*_i)_{ae} \delta_{bf}
\end{subeqnarray}
In order to diagonalize this matrix we consider the bilinear form, 
\begin{equation}
\mathcal{M}_F=\big(\hat{\Psi}^{\dag}_{R}\big)_{ab} M^{(R S)}_{ab, \, ef}
\big(\hat{\Psi}^T_{S}\big)_{ef}
\end{equation}
We expand the fermionic fields 
$\hat{\Psi}_R$ in fuzzy spherical harmonics as, 
\begin{equation}
\hat{\Psi}_R = \sum_{lm} \Psi^{(R)}_{lm} \hat{Y}_{lm}
\label{exp}
\end{equation}
This expansion then yields, 
\begin{equation}
\mathcal{M}_F= 4\eta^2
\sum^{N-1}_{l=0} \sum^{l}_{m=-l} \sum^{N-1}_{l'=0} \sum^{l'}_{m'=-l'} 
\big(\Psi^{\, (R)}_{lm}\big)^{\dag} \Psi^{\, (S)}_{l'm'} 
N^{(R S)}_{lm, \, l'm'}
\end{equation}
with 
\begin{equation}\label{eq:fermass}
N^{(R S)}_{lm, \, l'm'}= \delta_{ll'}
\left( \begin{array}{cccc}
J_{(L)}^{\, 2} + 1 & -iJ^{(L)}_3 & iJ^{(L)}_2 & 0 \\
iJ^{(L)}_3 & J_{(L)}^{\, 2} + 1 & -iJ^{(L)}_1 & 0 \\
-iJ^{(L)}_2 & iJ^{(L)}_1 & J_{(L)}^{\, 2} +1 & 0 \\
0 & 0 & 0 & J_{(L)}^{\, 2}
\end{array} \right)_{mm'}
\end{equation}
with $L=2l+1$.

The bosonic mass matrix receives a contribution from the scalar potential and 
a contribution from the covariant derivative of the complex scalars. The scalar
potential of the $\mathcal{N}=1^*$ theory is,
\begin{equation}
V = 4\eta^4 \textrm{Tr}_{N} \hat{H}^{\dag}_{ij} \hat{H}^{\phantom{\dag}}_{ij} 
+ 2 \eta^4 \textrm{Tr}_N \hat{D}^2
\end{equation}
where,
\begin{eqnarray*}
\hat{H}_{ij} & = & [\hat{\Phi}_i, \hat{\Phi}_j] - i\varepsilon_{ijk} \, 
\hat{\Phi}_k \\
& = & [J_i, \delta \hat{\Phi}_j] - [J_j, \delta \hat{\Phi}_i] 
+ [\delta \hat{\Phi}_i, \delta \hat{\Phi}_j] - i\varepsilon_{ijk} \, 
\delta \hat{\Phi}_k 
\end{eqnarray*}
\begin{eqnarray*}
\hat{D}^2 & = & [\hat{\Phi}^{\dag}_i, \hat{\Phi}^{\phantom{\dag}}_i]^2 \\
& = & \left( [J_i, \delta \hat{\Phi}^{\dag}_i] + [J_i, \delta 
\hat{\Phi}^{\phantom{\dag}}_i] + [\delta\hat{\Phi}^{\dag}_i, 
\delta\hat{\Phi}^{\phantom{\dag}}_i] \right)^2
\end{eqnarray*}
The scalar potential to quadratic order in the fields is,
\begin{equation}
\begin{split}
V = 2\eta^4 \textrm{Tr}_N \Big\lbrace & - 4[J_i, \delta 
\hat{\Phi}^{\dag}_j][J_i, \delta \hat{\Phi}^{\phantom{\dag}}_j] 
+ 4[J_i, \delta \hat{\Phi}^{\dag}_j][J_j, \delta \hat{\Phi}^{\phantom{\dag}}_i]
+ 8i \varepsilon_{ijk} \delta \hat{\Phi}^{\dag}_k [J_i, \delta 
\hat{\Phi}^{\phantom{\dag}}_j] \\
& + 4\delta \hat{\Phi}^{\dag}_i \delta \hat{\Phi}^{\phantom{\dag}}_i + 
[J_i, \delta\hat{\Phi}^{\dag}_i]^2 + 2[J_i, \delta\hat{\Phi}^{\dag}_i]
[J_j, \delta\hat{\Phi}^{\phantom{\dag}}_j] + 
[J_i, \delta\hat{\Phi}^{\phantom{\dag}}_i]^2 \Big\rbrace
\end{split}
\end{equation}
Using the cyclicity of the trace and imposing the gauge condition $[J_i, \delta \hat{\Phi}_i]=0$ the second term becomes,
\begin{equation}
\begin{split}
\textrm{Tr}_N [J_i, \delta \hat{\Phi}^{\dag}_j][J_j, \delta 
\hat{\Phi}^{\phantom{\dag}}_i] & = \textrm{Tr}_N \left( 
[\delta \hat{\Phi}^{\dag}_j, J_j][\delta \hat{\Phi}^{\phantom{\dag}}_i, J_i] 
- [\delta \hat{\Phi}^{\dag}_j, \delta \hat{\Phi}^{\phantom{\dag}}_i][J_j, J_i]
\right) \\
& = - \textrm{Tr}_N \, i \varepsilon_{ijk} \, \delta\hat{\Phi}^{\dag}_k 
[J_i, \delta \hat{\Phi}^{\phantom{\dag}}_j]
\end{split}
\end{equation}
Hence under the gauge condition the scalar potential is reduced to,
\begin{equation}
V = 8\eta^4 \textrm{Tr}_N \Big\lbrace \delta \hat{\Phi}^{\dag}_j [J_i, [J_i, 
\delta \hat{\Phi}^{\phantom{\dag}}_j]] 
+ i\varepsilon_{ijk} \, \delta \hat{\Phi}^{\dag}_i [J_j, \delta 
\hat{\Phi}^{\phantom{\dag}}_k] + \delta \hat{\Phi}^{\dag}_i \delta 
\hat{\Phi}^{\phantom{\dag}}_i \Big\rbrace
\end{equation}
The bosonic mass contribution is,
\begin{equation}
\mathcal{M}_V = 4\eta^2 \textrm{Tr}_N \Big\lbrace \delta \hat{\Phi}^{\dag}_j [J_i, 
[J_i, \delta \hat{\Phi}^{\phantom{\dag}}_j]] 
+ i\varepsilon_{ijk} \, \delta \hat{\Phi}^{\dag}_i [J_j, \delta 
\hat{\Phi}^{\phantom{\dag}}_k] + \delta \hat{\Phi}^{\dag}_i \delta 
\hat{\Phi}^{\phantom{\dag}}_i \Big\rbrace
\end{equation}
Expanding perturbations $\delta \hat{\Phi}_i$ in fuzzy spherical harmonics,
\begin{subeqnarray}
\delta \hat{\Phi}^{\phantom{\dag}}_i & = & \sum_{l,m} \phi^{(i)}_{lm} 
\hat{Y}^{\phantom{\dag}}_{lm} \\
\delta \hat{\Phi}^{\dag}_i & = & \sum_{l,m} \bar{\phi}^{(i)}_{lm} 
\hat{Y}^{\dag}_{lm}
\end{subeqnarray}
then,
\begin{equation}
\mathcal{M}_V = 4 \eta^2 \sum_{l,m,l',m'} \big(\phi^{\, i}_{lm}\big)^{\dag} 
\phi^{\, k}_{l'm'} N^{(ik)}_{lm,l'm'}
\end{equation}
where, 
\begin{equation}
N^{(ik)}_{lm,l'm'} = \left( J^2_{(L)} + 1 \right)_{mm'} \delta_{ll'} 
\delta_{ik} + i\varepsilon_{ijk} \left( J^{(L)}_j \right)_{mm'} \delta_{ll'}
\end{equation}

The covariant derivative of the complex scalar is,
\begin{equation}
2 \eta^2 \textrm{Tr}_{N} D_{\mu} \hat{\Phi}^{\dag}_i D^{\mu} 
\hat{\Phi}^{\phantom{\dag}}_i 
= 2 \eta^2 \textrm{Tr}_{N} \left( \partial_{\mu} \hat{\Phi}^{\dag}_i 
+ i\eta[A_{\mu}, \hat{\Phi}^{\dag}_i] \right) \left( \partial^{\mu} 
\hat{\Phi}^{\phantom{\dag}}_i + i\eta[A^{\mu}, \hat{\Phi}^{\phantom{\dag}}_i]
\right)
\end{equation}
This contributes to the bosonic mass matrix,
\begin{equation}
\begin{split}
\mathcal{M}_D = - 4 \eta^2 \, \textrm{Tr}_{N} [J_i, \hat{A}_{\mu}][J_i, \hat{A}^{\mu}]
& = 4\eta^2 \, \textrm{Tr}_N \hat{A}_{\mu} [J^2, \hat{A}^{\mu} ] \\
& = 4\eta^2 \sum_{lm,l'm'} a_{(\mu) \, lm} a^{(\mu)}_{l'm'} \left( 
J^2_{(L)} \right)_{mm'} \delta_{ll'}
\end{split}
\end{equation}
The Bosonic Mass Matrix is,
\begin{equation}
\mathcal{M}_B = \big(\hat{\Phi}^{\dag}_R \big)_{ab} M^{(RS)}_{ab, \, ef}
\big(\hat{\Phi}^T_S\big)_{ef} = 4\eta^2 \sum^{N-1}_{l=0} \sum^{l}_{m=-l} 
\sum^{N-1}_{l'=0} \sum^{l'}_{m'=-l'} \big(\phi^{\, R}_{lm}\big)^{\dag} 
\phi^{\, S}_{l'm'} 
N^{(RS)}_{lm, \, l'm'}
\end{equation}
where the three complex scalars and gauge boson are combined in a column vector
$\hat{\Phi}_R$ with $\hat{\Phi}_R = \hat{\Phi}_i$ for $i=1,2,3$ and $\hat{\Phi}_4 = 
\hat{A}_{\mu}
$. The matrix $N^{(RS)}_{lm, \, l'm'}$ is the same matrix as (\ref{eq:fermass}).

The fermionic and bosonic calculations lead to the same matrix. To complete the
diagonalization we must solve the characteristic equation of the matrix 
$N^{(RS)}_{lm, \, l'm'}$. Consider the $(p+q)\times(p+q)$ matrix,
\begin{equation}
{\cal X} = \begin{array}{cc}
& \begin{array}{cc} (p) & (q) \end{array} \\
\begin{array}{c} (p) \\ (q) \end{array} &
\left(\begin{array}{cc} 
A & B \\ C & D 
\end{array} \right)
\end{array}
\end{equation}
The determinant of $\mathcal{X}$ can be evaluated using the formula,
\begin{equation}
Det(\mathcal{X}) = Det(A)Det(D-CA^{-1}B)
\end{equation}
Clearly in the mass matrix $N^{(RS)}_{lm, \, l'm'}$ the gauge boson/gaugino 
contribution is trivial and we will concentrate on calculating the determinant 
of,
\begin{equation}
\tilde{N}_{m m'} = \left( \begin{array}{ccc}
\gamma^{(L)}.\mathbb{1} & -iJ^{(L)}_3 & iJ^{(L)}_2 \\
iJ^{(L)}_3 & \gamma^{(L)}.\mathbb{1} & -iJ^{(L)}_1 \\
-iJ^{(L)}_2 & iJ^{(L)}_1 & \gamma^{(L)}.\mathbb{1}
\end{array} \right)_{mm'}
\end{equation}
where $\gamma^{(L)} = l(l+1) + 1 - \lambda$ and $\lambda$ is the eigenvalue of 
the characteristic equation,
\begin{equation}
\textrm{Det} \left( N - \lambda \mathbb{1} \right) = 0
\end{equation}
We define the matrix,
\begin{equation}
A = \left( \begin{array}{cc}
\gamma^{(L)}.\mathbb{1} & -i J^{(L)}_3 \\
i J^{(L)}_3 & \gamma^{(L)}.\mathbb{1}
\end{array} \right)
\end{equation}
whose inverse is,
\begin{equation}
A^{-1} =
\left( \begin{array}{cc}
a & b \\ -b & a
\end{array} \right)
\end{equation}
where,
\begin{subeqnarray}
a & = & \frac{\gamma^{(L)}}{(\gamma^{(l)})^2 - m^2} \, \delta_{mm'} \\
b & = & \frac{im}{(\gamma^{(L)})^2 - m^2} \, \delta_{mm'}
\end{subeqnarray}
We find that,
\begin{equation}
Det(N - \lambda \mathbb{1}) = \prod^{N-1}_{l=0} 
\prod^l_{m=-l} (\gamma^{(L)}-1)^{2(2l+1)} (\gamma^{(L)}+l)^{2l+3} 
\big(\gamma^{(L)}-(l+1)\big)^{2l-1}
\end{equation}
The roots of this characteristic equation yield the eigenvalues of the mass 
matrices.

For $l=0$ we therefore find, 
\begin{center}
\begin{tabular}{c c}
\hline
Eigenvalue & Degeneracy \\
\hline
$0$   & $1$ \\
$1$ & $3$ \\
\hline
\end{tabular}
\end{center}
while for $l=1,2,\ldots,N-1$ we get, 
\begin{center}
\begin{tabular}{c c}
\hline
Eigenvalue & Degeneracy \\
\hline
$l^{2}$   & $2l-1$ \\
$l(l+1)$ & $2(2l+1)$ \\
$(l+1)^{2}$   & $2l+3$ \\
\hline
\end{tabular}
\end{center}
To find the complete spectrum in this case 
we sum over all values of $l$. The final result is a single zero 
eigenvalue and two series 
of eigenvalues labeled by a positive integer $k=1,2,\ldots N-1$, 
\begin{center}
\begin{tabular}{c c}
\hline
Eigenvalue & Degeneracy \\
\hline
$k^{2}$   & $4k$ \\
$k(k+1)$ & $2(2k+1)$ \\
\hline
\end{tabular}
\end{center}
Finally we find one extra eigenvalue $\lambda=N^{2}$ with 
degeneracy $2N+1$.

Each eigenvalue of the fermionic mass matrix corresponds to a single 
left-handed Weyl fermion and it's right-handed charge conjugate. Similarly each
eigenvalue of the bosonic matrix corresponds to a complex scalar or gauge 
boson. The theory has $\mathcal{N}=1$ supersymmetry, therefore all the states 
must 
form $\mathcal{N}=1$ supersymmetry multiplets. We began with a $U(N)$ gauge 
group which is broken to $U(1)$. The spectrum must contain a single massless 
gauge boson and $N^2-1$ massive gauge bosons. Clearly the $l=0$ state must form
a massless vector multiplet. The $N^2-1$ massive gauge bosons must form $N^2-1$
massive vector multiplets, which are formed from a massless vector multiplet 
and a massless chiral multiplet. The remaining states must form massive chiral
multiplets.

In addition to a single massless vector multiplet 
we have two towers of multiplets labelled by $k=1,2,\ldots, N-1$ as
tabulated below,
\begin{center}  
\begin{tabular}{c c l}
\hline
$M^2$ & Degeneracy & Multiplet \\
\hline
$\eta^2 k(k+1)$ & $2k+1$ & Massive Vector \\
$\eta^2 k^{2}$   & $4k$  & Massive Chiral \\
\hline
\end{tabular}
\end{center} 
The spectrum is completed by $2N+1$ chiral multiplets of mass 
$M^2 = \eta^2 N^{2}$. In the limit $N\rightarrow \infty$ this precisely matches the 
spectrum of the Maldacena-N\'u\~nez compactification with the identification $\eta 
\sim \frac{1}{R}$, up to a numerical coefficient. For finite $N$ 
the spectrum of the ${\cal N}=1^{*}$ theory is a subset of that of the 
six-dimensional theory obtained by retaining only those states with
mass less than $N^{2}$ (together with $2N+1$ chiral multiplets of mass
equal to $N^{2}$).
\paragraph{}
We can generalise the above calculation to the more 
general Higgs vacua discussed in section \ref{sec:decon}, where the 
gauge group is broken from $U(N) \to U(p)$ with $N=pq$. 
In this case, the vacuum expectation value takes
the form,
\begin{equation}
<\Phi_i> = \mathbb{1}_{(p)} \otimes J^{(q)}_i
\end{equation}
In this vacuum the field fluctuations decompose as a tensor product,
\begin{equation}
\delta\hat{\Phi}_i = \sum_{l,m} \phi^{(i)}_{lm(p)} \otimes \hat{Y}^{(q)}_{lm}
\end{equation}
The $q^2$ coefficients $\phi^{(i)}_{lm}$ are $p \times p$ matrices and the 
fuzzy spherical harmonics are $q \times q$ matrices. 
The $k=0$ mode remains a massless vector multiplet, but now has degeneracy 
$p^2$. Note that, in constrast to the maximally Higgsed vacuum, the
$U(p)$ low-energy effective theory is asymptotically free and a classical
analysis is only reliable at scales much higher than the corresponding
dynamical scale. 
\paragraph{}
The remaining $q^2-1$ modes (at finite $N$, $q$) are,
\begin{center}  
\begin{tabular}{c c l}
\hline
$M^2$ & Degeneracy & Multiplet \\
\hline
$\eta^2 k(k+1)$ & $(2k+1)p^2$ & Massive Vector \\
$\eta^2k^{2}$   & $4k \, p^2$  & Massive Chiral \\
\hline
\end{tabular}
\end{center} 
for $k=1,2,\ldots, q-1$. The spectrum is completed by $(2q+1)p^2$ 
extra massive chiral multiplets with mass $M^2=\eta^2 q^{2}$. 
Notice that the degeneracies of all states are integer multiples of
$p^{2}$ consistent with each state transforming in the adjoint of the
unbroken $U(p)$ gauge symmetry. Finally, we 
can also study the spectrum corresponding to vacua of the 
$SU(N)$ theory. The effect is simply to replace $p^{2}$ by $p^{2}-1$
appropriate for adjoint multiplets of an unbroken $SU(p)$. 
\paragraph{}
In each of the cases discussed above, 
the spectrum matches the Kaluza-Klein spectrum of a six-dimensional
theory with non-abelian gauge group ($U(p)$ for the $U(N)$ theory and
$SU(p)$ for $SU(N)$). In fact this spectrum can be obtained from a
MN twisted compactification of the corresponding ${\cal N}=(1,1)$ SUSY 
Yang-Mills theory by an easy generalisation of the analysis of Section
\ref{sec:MN}.

\section{Effective Six-Dimensional Theory}\label{sec:eff}

In the $N \to \infty$ limit, the classical spectrum 
of the $\mathcal{N}=1^*$ theory is identical to that 
of the MN compactification. We now calculate the effective 
action of the Higgsed $\mathcal{N}=1^*$ theory and compare the 
action with the six-dimensional bosonic action 
(\ref{result}) action derived in section \ref{sec:MNA}. We follow the 
deconstruction procedure described in section \ref{sec:decon}, mapping the 
four-dimensional matrix model of the Higgsed $\mathcal{N}=1^*$ theory to a 
six-dimensional non-commutative field theory, and then take the commutative 
limit ($N \to \infty$). We will break up the calculation into four parts.
We begin with the scalar potential, with gauge condition $[J_i, 
\delta\hat{\Phi}_i] = 0$ imposed.
\begin{eqnarray*}
V & = &  4 \eta^4 \textrm{Tr}_{N} \hat{H}^{\dag}_{ij} \hat{H}^{\phantom{\dag}}
_{ij} + 2\eta^4 \textrm{Tr}_N \hat{D}^2 \\
& = & 8 \eta^4 \textrm{Tr}_{N} \Big\lbrace \delta \hat{\Phi}^{\dag}_j [J_i, 
[J_i, \delta \hat{\Phi}^{\phantom{\dag}}_j]] 
+ i\varepsilon_{ijk} \delta \hat{\Phi}^{\dag}_i [J_j, \delta \hat{\Phi}
^{\phantom{\dag}}_k] + \delta \hat{\Phi}^{\dag}_i \delta \hat{\Phi}
^{\phantom{\dag}}_i \\
\nonumber && - [J_i, \delta \hat{\Phi}^{\phantom{\dag}}_j][\delta \hat{\Phi}
^{\dag}_j, \delta \hat{\Phi}^{\phantom{\dag}}_i] + [J_i, \delta \hat{\Phi}
^{\dag}_j][\delta \hat{\Phi}^{\dag}_i, \delta \hat{\Phi}^{\phantom{\dag}}_j] 
+ i\varepsilon_{ijk}[\delta \hat{\Phi}^{\dag}_i, \delta \hat{\Phi}^{\dag}_j]
\delta \hat{\Phi}^{\phantom{\dag}}_k \\
\nonumber && + i\varepsilon_{ijk}\delta \hat{\Phi}^{\dag}_i 
[\delta \hat{\Phi}^{\phantom{\dag}}_j, \delta \hat{\Phi}^{\phantom{\dag}}_k]
- [\delta \hat{\Phi}^{\dag}_i, \delta \hat{\Phi}^{\dag}_j][\delta \hat{\Phi}
^{\phantom{\dag}}_i, \delta \hat{\Phi}^{\phantom{\dag}}_j] + \frac{1}{4} 
[\delta\hat{\Phi}^{\dag}_i, \delta\hat{\Phi}^{\phantom{\dag}}_i]^2 \Big\rbrace
\end{eqnarray*}
where,
\begin{eqnarray*}
\hat{H}_{ij} & = & [\hat{\Phi}_i, \hat{\Phi}_j] - i\varepsilon_{ijk} \, 
\hat{\Phi}_k \\
& = & [J_i, \delta \hat{\Phi}_j] - [J_j, \delta \hat{\Phi}_i] + [\delta 
\hat{\Phi}_i, \delta \hat{\Phi}_j] - i\varepsilon_{ijk} \, \delta \hat{\Phi}_k 
\\
\hat{D} & = & [J_i, \delta \hat{\Phi}^{\dag}_i] + [J_i, \delta 
\hat{\Phi}^{\phantom{\dag}}_i] + [\delta\hat{\Phi}^{\dag}_i, 
\delta\hat{\Phi}^{\phantom{\dag}}_i]
\end{eqnarray*}
From the correspondence between matrices and functions (equations \eqref{eq:map}
and \eqref{eq:maptr}),
\begin{eqnarray*}
V & = & 8 \frac{N}{4 \pi} \, \eta^4 \int d\Omega \bigg\lbrace \delta 
\Phi^{\dag}_i \left(L^2 \, \delta \Phi^{\phantom{\dag}}_i \right) 
+ i\varepsilon_{ijk} \delta \Phi^{\dag}_i \left( L_j \, \delta \Phi
^{\phantom{\dag}}_k \right) + \delta \Phi^{\dag}_i \delta \Phi
^{\phantom{\dag}}_i \\
\nonumber && - (L_i \delta \Phi^{\phantom{\dag}}_j)[\delta \Phi^{\dag}_j, 
\delta \Phi^{\phantom{\dag}}_i] + (L_i \delta \Phi^{\dag}_j)[\delta \Phi
^{\dag}_i, \delta \Phi^{\phantom{\dag}}_j] + i\varepsilon_{ijk}[\delta \Phi
^{\dag}_i, \delta \Phi^{\dag}_j]\delta \Phi^{\phantom{\dag}}_k \\
\nonumber && + i\varepsilon_{ijk}\delta \Phi^{\dag}_i [\delta \Phi
^{\phantom{\dag}}_j, \delta \Phi^{\phantom{\dag}}_k] - [\delta \Phi^{\dag}_i,
\delta \Phi^{\dag}_j][\delta \Phi^{\phantom{\dag}}_i, \delta \Phi
^{\phantom{\dag}}_j] + \frac{1}{4} [\delta\Phi^{\dag}_i, \delta\Phi
^{\phantom{\dag}}_i]^2 \bigg\rbrace_{\star}
\end{eqnarray*}
where $\lbrace \rbrace_{\star}$ means all products are non-commutative 
star-products. Taking the commutative limit \footnote{Note commutators of field
fluctuations vanish in this limit, see equation \eqref{eq:nocom}, e.g. 
$[\delta\Phi_i, \delta\Phi_j] = 0$},
\begin{equation}
V = 8 \frac{N}{4 \pi} \, \eta^4 \int d\Omega \bigg\lbrace \delta \Phi^{\dag}_i
\left(L^2 \, \delta \Phi^{\phantom{\dag}}_i \right) 
+ i\varepsilon_{ijk} \delta \Phi^{\dag}_i \left( L_j \, \delta \Phi
^{\phantom{\dag}}_k \right) + \delta \Phi^{\dag}_i \delta \Phi
^{\phantom{\dag}}_i \bigg\rbrace 
\end{equation}
The bosonic kinetic terms are,
\begin{eqnarray*}
\mathcal{L}_{Bkin} &=& \eta^2 \textrm{Tr}_N \left\lbrace -\frac{1}{4} \hat{F}
_{\mu \nu} \hat{F}^{\mu \nu} - 2 D_{\mu} \hat{\Phi}^\dag _i D^{\mu} \hat{\Phi}
_i \right\rbrace \\
\nonumber & = & \eta^2 \textrm{Tr}_N \Big\lbrace -\frac{1}{4} \hat{F}_{\mu \nu}
\hat{F}^{\mu \nu} - 2\Big( \partial_{\mu} \delta\hat{\Phi}^\dag _i \partial
^{\mu} \delta\hat{\Phi}^{\phantom{\dag}}_i - i\eta[J_i,\hat{A}_{\mu}]
\partial^{\mu}\delta\hat{\Phi}^{\phantom{\dag}}_i \\
&& + i\eta[\hat{A}_{\mu}, \delta\hat{\Phi}^{\dag}_i] \partial^{\mu}
\delta\hat{\Phi}^{\phantom{\dag}}_i - 
i\eta \partial_{\mu} \delta\hat{\Phi}^{\dag}_i [J_i, \hat{A}^{\mu}] + i \eta
\partial_{\mu} \delta\hat{\Phi}^{\dag}_i [ \hat{A}^{\mu}, \delta\hat{\Phi}
^{\phantom{\dag}}_i] \\
\nonumber && - \eta^2 [J_i,\hat{A}_{\mu}][J_i, \hat{A}^{\mu}] + \eta^2 [J_i, 
\hat{A}_{\mu}]
[ \hat{A}^{\mu}, \delta\hat{\Phi}^{\phantom{\dag}}_i] + \eta^2 [\hat{A}_{\mu},
\delta\hat{\Phi}^{\dag}_i][J_i, \hat{A}^{\mu}] \\
&& - \eta^2 [\hat{A}_{\mu},\delta\hat{\Phi}^{\dag}_i][ \hat{A}^{\mu}, 
\delta\hat{\Phi}^{\phantom{\dag}}_i] \Big)\Big\rbrace
\end{eqnarray*}
where $\hat{F}_{\mu \nu} = \partial_{\mu} \hat{A}_{\nu} - \partial_{\nu} 
\hat{A}_{\mu} + i\eta [\hat{A}_{\mu}, \hat{A}_{\nu}]$ and $D_{\mu} \hat{\Phi}_i
= \partial_{\mu} \hat{\Phi}_i + i\eta[\hat{A}_{\mu}, \hat{\Phi}_i]$.
From the correspondence between matrices and functions,
\begin{eqnarray*}
\mathcal{L}_{Bkin} & = & \frac{N}{4 \pi} \, \eta^2 \int d\Omega \Big\lbrace 
-\frac{1}{4} F_{\mu \nu} F^{\mu \nu}  - 2\Big( \partial_{\mu} (\delta \Phi
^{\dag}_i) \partial^{\mu} (\delta \Phi^{\phantom{\dag}}_i) - 
i\eta\partial_{\mu} (\delta \Phi^{\dag}_i) (L_i A^{\mu})\\
&& + i\eta\partial_{\mu} (\delta \Phi^{\dag}_i)[A^{\mu}, \delta\Phi
^{\phantom{\dag}}_i] - i\eta(L_i A_{\mu}) \partial^{\mu} (\delta \Phi
^{\phantom{\dag}}_i) + i\eta[A_{\mu}, \delta\Phi^{\dag}_i]\partial^{\mu}
(\delta \Phi^{\phantom{\dag}}_i) \\
&& - \eta^2 (L_i A_{\mu})(L_i A^{\mu})  + \eta^2 (L_i A_{\mu})[A^{\mu}, 
\delta\Phi^{\phantom{\dag}}_i] + \eta^2 [A_{\mu}, \delta\Phi^{\dag}_i]
(L_i A^{\mu}) \\
&& + \eta^2 [A_{\mu}, \delta\Phi^{\dag}_i][A^{\mu}, 
\delta\Phi^{\phantom{\dag}}_i]\Big) \Big\rbrace_{\star} 
\end{eqnarray*}
Taking the commutative limit and applying the gauge condition $L_i \delta
\Phi_i = 0$,
\begin{equation}
\mathcal{L}_{Bkin} = \frac{N}{4 \pi} \, \eta^2 \int d\Omega \Big\lbrace 
-\frac{1}{4} F_{\mu \nu} F^{\mu \nu}  - 2 \partial_{\mu} (\delta \Phi
^{\dag}_i) \partial^{\mu} (\delta \Phi^{\phantom{\dag}}_i) 
+ 2\eta^2(L_i A_{\mu})(L_i A^{\mu}) \Big\rbrace
\end{equation}
where now $F_{\mu \nu} = \partial_{\mu} A_{\nu} - \partial_{\nu} A_{\mu}$.
Consider the fermionic interaction terms,
\begin{eqnarray*}
\mathcal{L}_{Fint} & = & \eta^4 \textrm{Tr}_N \Big\lbrace i\hat{\psi}_i 
\epsilon_{ijk}[\hat{\Phi}_k , \hat{\psi}_j] + i \hat{\bar{\psi}}_i 
\epsilon_{ijk}[\hat{\Phi}^{\dag}_k , \hat{\bar{\psi}}_j] - 
i\hat{\lambda} [ \hat{\Phi}^{\dag}_i , \hat{\psi}_i ] \\ 
\nonumber & & + i\hat{\psi}_i [\hat{\Phi}^{\dag}_i , \hat{\lambda} ] + i
\hat{\bar{\psi}}_i [ \hat{\Phi}_i , \hat{\bar{\lambda}} ] - 
i\hat{\bar{\lambda}} [ \hat{\Phi}_i , \hat{\bar{\psi}}_i ] - \hat{\psi}_i 
\hat{\psi}_i -  \hat{\bar{\psi}}_i \hat{\bar{\psi}}_i \Big\rbrace \\
& = & \eta^4 \textrm{Tr}_N \Big\lbrace i\hat{\psi}_i \epsilon_{ijk}[J_k , 
\hat{\psi}_j] + i\hat{\psi}_i \epsilon_{ijk}[\delta\hat{\Phi}_k , 
\hat{\psi}_j] + i \hat{\bar{\psi}}_i \epsilon_{ijk}[J_k , 
\hat{\bar{\psi}}_j] \\ 
& & + i \hat{\bar{\psi}}_i \epsilon_{ijk}[\delta\hat{\Phi}^{\dag}_k , 
\hat{\bar{\psi}}_j] - i\hat{\lambda} [ J_i , \hat{\psi}_i ] - i\hat{\lambda} 
[ \delta\hat{\Phi}^{\dag}_i , \hat{\psi}_i ] + i\hat{\psi}_i 
[J_i , \hat{\lambda} ] + i\hat{\psi}_i [\delta\hat{\Phi}^{\dag}_i , 
\hat{\lambda} ] \\
& & \nonumber + i\hat{\bar{\psi}}_i [ J_i , \hat{\bar{\lambda}} ] + 
i\hat{\bar{\psi}}_i [ \delta\hat{\Phi}_i , \hat{\bar{\lambda}} ] - 
i\hat{\bar{\lambda}} [ J_i , \hat{\bar{\psi}}_i ] - i\hat{\bar{\lambda}} 
[ \delta\hat{\Phi}_i , \hat{\bar{\psi}}_i ] - \hat{\psi}_i \hat{\psi}_i - 
\hat{\bar{\psi}}_i \hat{\bar{\psi}}_i
\Big\rbrace
\end{eqnarray*}
From the correspondence between matrices and functions,
\begin{eqnarray*}
\nonumber \mathcal{L}_{Fint} & = & \frac{N}{4 \pi} \, \eta^4  \int d\Omega 
\Big\lbrace i\psi_i \epsilon_{ijk}\left(L_k \psi_j \right) + i\psi_i 
\epsilon_{ijk}[\delta\Phi_k , \psi_j] + i \bar{\psi}_i \epsilon_{ijk}\left(L_k 
\bar{\psi}_j \right) \\ 
& & + i \bar{\psi}_i \epsilon_{ijk}[\delta\Phi^{\dag}_k , \bar{\psi}_j] - 
i\lambda \left(L_i \psi_i \right)- i\lambda [ \delta\Phi^{\dag}_i , \psi_i ] + 
i\psi_i \left(L_i \lambda \right) + i\psi_i [\delta\Phi^{\dag}_i , \lambda ] \\
& & \nonumber + i\bar{\psi}_i \left(L_i \bar{\lambda} \right) + i\bar{\psi}_i 
[ \delta\Phi_i, \bar{\lambda} ] - i\bar{\lambda} \left(L_i \bar{\psi}_i \right) 
- i\bar{\lambda} [ \delta\Phi_i , \bar{\psi}_i ] - \psi_i \psi_i - \bar{\psi}_i
\bar{\psi}_i
\Big\rbrace_{\star}
\end{eqnarray*}
In the commutative limit,
\begin{equation}
\begin{split}
\mathcal{L}_{Fint} =  \frac{N}{4 \pi} \, \eta^4 \int d\Omega \Big\lbrace &
i\psi_i \,  \varepsilon_{ijk} L_k \psi_j  + i \bar{\psi}_i \, \varepsilon_{ijk}
L_k \bar{\psi}_j  - i\lambda L_i \psi_i \\
& + i\psi_i L_i \lambda  + i\bar{\psi}_i L_i \bar{\lambda} -
i\bar{\lambda} L_i \bar{\psi}_i  - \psi_i \psi_i - \bar{\psi}_i \bar{\psi}_i
\Big\rbrace
\end{split}
\end{equation}
The fermionic kinetic terms are,
\begin{eqnarray*}
\mathcal{L}_{Fkin} & = & -i \eta^3 \textrm{Tr}_N \Big( \hat{\lambda} 
\sigma^{\mu} D_{\mu} \hat{\bar{\lambda}} + \hat{\psi}_i \sigma^{\mu} D_{\mu} 
\hat{\bar{\psi}}_i \Big) \\
& = & -i \eta^3 \textrm{Tr}_N \Big( \hat{\lambda} \sigma^{\mu} \partial_{\mu} 
\hat{\bar{\lambda}} + i\eta \hat{\lambda} \sigma^{\mu} [\hat{A}_{\mu}, 
\hat{\bar{\lambda}}] + \hat{\psi}_i \sigma^{\mu} \partial_{\mu} 
\hat{\bar{\psi}}_i + i\eta \hat{\psi}_i \sigma^{\mu} [\hat{A}_{\mu}, 
\hat{\bar{\psi}}_i] \Big)
\end{eqnarray*}
From the correspondence between matrices and functions,
\begin{equation*}
\begin{split}
\mathcal{L}_{Fkin} = -i \frac{N}{4 \pi} \, \eta^3 \int d\Omega \Big( & \lambda 
\sigma^{\mu} \partial_{\mu} \bar{\lambda} + i\eta \lambda \sigma^{\mu} [A_{\mu}, 
\bar{\lambda}] + \psi_i \sigma^{\mu} \partial_{\mu} \bar{\psi}_i \\
& + i\eta \psi_i 
\sigma^{\mu} [A_{\mu}, \bar{\psi}_i] \Big)_{\star}
\end{split}
\end{equation*}
In the commutative limit,
\begin{equation}
\mathcal{L}_{Fkin} = -i \frac{N}{4 \pi} \, \eta^3 \int d\Omega \Big( \lambda 
\sigma^{\mu} \partial_{\mu} \bar{\lambda} + \psi_i \sigma^{\mu} \partial_{\mu} 
\bar{\psi}_i \Big)
\end{equation}
Summarising the action,
\begin{equation}
\begin{split}
\mathcal{S} = & \frac{1}{g^2_{ym}} \frac{N}{4 \pi} \, \eta^2 \int 
d^4 x \, \int d\Omega \bigg\lbrace -\frac{1}{4} F_{\mu \nu} F^{\mu \nu}  - 2 
\partial_{\mu} (\delta \Phi^{\dag}_i) \partial^{\mu} (\delta \Phi
^{\phantom{\dag}}_i) \\
& - i \eta \lambda \sigma^{\mu} \partial_{\mu} 
\bar{\lambda} - i \eta \psi_i \sigma^{\mu} \partial_{\mu} \bar{\psi}_i + \eta^2 
\Big( 2(L_i A_{\mu})(L_i A^{\mu}) + i\psi_i \varepsilon_{ijk} \, L_k \psi_j  \\
&  + i \bar{\psi}_i \varepsilon_{ijk} \, L_k \bar{\psi}_j  - i\lambda 
L_i \psi_i + i\psi_i L_i \lambda  + i\bar{\psi}_i L_i \bar{\lambda} - 
i\bar{\lambda} L_i \bar{\psi}_i  - \psi_i \psi_i \\
& - \bar{\psi}_i \bar{\psi}_i - 8\delta \Phi^{\dag}_i \left(L^2 \, 
\delta \Phi^{\phantom{\dag}}_i \right)  - 8i \varepsilon_{ijk} \, \delta \Phi
^{\dag}_i \left( L_j \, \delta \Phi^{\phantom{\dag}}_k \right) - 8\delta \Phi
^{\dag}_i \delta \Phi^{\phantom{\dag}}_i \Big) \bigg\rbrace
\end{split}
\end{equation}
Before proceeding further we want to rewrite this action with Majorana spinors.
 We define the Majorana spinors,
\begin{equation}
\Psi_{i A} = \left( \begin{array}{c} \psi_{i \alpha} \\ \bar{\psi}^{\phantom{i} 
\dot{\alpha}}_i \end{array} \right) \qquad \qquad \Lambda_A = \left( 
\begin{array}{c} \lambda_{\alpha} \\ \bar{\lambda}^{\dot{\alpha}}_i 
\end{array} \right)
\end{equation}
with $SO(3,1)$ index $A$.
The action is now,
\begin{equation}
\begin{split}
\mathcal{S} = & \frac{1}{g^2_{ym}} \frac{N}{4 \pi} \, \eta^2 \int 
d^4 x \, \int d\Omega \bigg\lbrace -\frac{1}{4} F_{\mu \nu} F^{\mu \nu}  - 2 
\partial_{\mu} (\delta \Phi^{\dag}_i) \partial^{\mu} (\delta 
\Phi^{\phantom{\dag}}_i) - \frac{i}{2} \, \eta \bar{\Lambda} \gamma^{\mu} 
\partial_{\mu} \Lambda \\
& - \frac{i}{2} \, \eta \bar{\Psi}_i \gamma^{\mu} \partial_{\mu}
\Psi_i + \eta^2 \Big( 2(L_i A_{\mu})(L_i A^{\mu}) + i \bar{\Psi}_i 
\varepsilon_{ijk} \, L_k \Psi_j + 2i \bar{\Psi}_i L_i \Lambda \\
& - \bar{\Psi}_i \Psi_i- 8\delta 
\Phi^{\dag}_i \left(L^2 \, \delta \Phi^{\phantom{\dag}}_i \right) - 8i 
\varepsilon_{ijk} \, \delta \Phi^{\dag}_i \left( L_j \, \delta \Phi
^{\phantom{\dag}}_k \right) - 8\delta \Phi^{\dag}_i \delta \Phi
^{\phantom{\dag}}_i \Big) \bigg\rbrace
\end{split}
\end{equation}

\paragraph{}

The action remains in the form inherited from the four-dimensional theory, it 
does not have the canonical form of a six-dimensional theory. Our calculation 
of the Higgsed $\mathcal{N}=1^*$ mass spectrum suggests that the eigenstates 
of the Higgsed $\mathcal{N}=1^*$ theory are the eigenstates of the sphere. We 
now perform a similarity transformation on these $\mathcal{N}=1^*$ fields and 
expand them in fields on the sphere. Our action will take the canonical form 
of a six-dimensional field theory.
We begin with the scalar potential. 
\begin{eqnarray}
V & = & 8 \eta^4 \int d\Omega \, \delta\Phi^{\dag}_i \Delta_{ij} \delta\Phi
^{\phantom{\dag}}_j \\
\Delta_{ij} & = & (L^2 + 1) \delta_{ij} - i\varepsilon_{ijk} L_k
\end{eqnarray}
We can express the complex scalars as $\delta\Phi_i = \frac{1}{\sqrt{2}} 
\left(a_i + ib_i \right)$ ($a_i$ and 
$b_i$ real) and define, 
\begin{equation}
\mathcal{Y}^{\hat{\alpha}}_i = \frac{1}{\sqrt{2}} \left(\begin{array}{c} a_i \\ 
i b_i \end{array}
\right)
\end{equation}
where the index $\hat{\alpha}$ labels the two components. The scalar potential is,
\begin{eqnarray*}
V & = & 8 \eta^4 \int d\Omega \, \mathcal{Y}^{\dag}_{i \hat{\alpha}} 
(\hat{O}_{ij})^{\hat{\alpha}}_{\phantom{\alpha} \hat{\beta}} \mathcal{Y}
^{\hat{\beta}}_j
\end{eqnarray*}
where the matrix $(\hat{O}_{ij})^{\hat{\alpha}}_{\phantom{\alpha} \hat{\beta}} = \delta^{\hat{\alpha}}_{\phantom{\alpha} \hat{\beta}} \Delta_{ij}$. We use our 
knowledge of the eigenstates of the 2-sphere to find the eigenvectors of the 
operator $\hat{O}_{ij}$. The complete set of eigenvectors of the operator 
$\hat{O}_{ij}$ are,
\begin{subeqnarray}
e^{\hat{\alpha}}_i & = & v^{\hat{\alpha}} \frac{1}{\sqrt{l(l+1)}} \, i L_i Y_{lm}
(\theta, \phi) \\
\chi^{\hat{\alpha}}_i & = & (\sigma_i)^{\hat{\alpha}}_{\phantom{\alpha} 
\hat{\beta}} \Omega^{\hat{\beta}}_{q_{\pm} lm}(\theta, \phi) + 
\frac{1}{\kappa_{\pm}} \, L_i \Omega^{\hat{\alpha}}_{q_{\pm} lm}(\theta, \phi)
\end{subeqnarray}
where $v^{\hat{\alpha}}$ is an arbitrary 2-component object.

We expand $\mathcal{Y}^{\hat{\alpha}}_i$ in the complete basis of eigenvectors,
\begin{equation}
\mathcal{Y}^{\hat{\alpha}}_i = \mathcal{A}^{\hat{\alpha}}_i + \mathcal{P}
^{\hat{\alpha}}_i
\end{equation}
where,
\begin{subeqnarray}
\mathcal{A}^{\hat{\alpha}}_i & = & \sum_{lm} v^{\hat{\alpha}}_{lm} 
\frac{1}{\sqrt{l(l+1)}} \, i L_i Y_{lm}(\theta, \phi) \\
\mathcal{P}^{\hat{\alpha}}_i & = & \sum_{lm} \left\lbrace \xi^+_{lm} \left( 
(\sigma_i)^{\hat{\alpha}}_{\phantom{\alpha} \hat{\beta}} 
\Omega^{\hat{\beta}}_{q_+ lm} + \frac{1}{\kappa_+} \, L_i \Omega^{\hat{\alpha}}
_{q_+ lm} \right) \right.\\
& & \nonumber \qquad \left. + \, \xi^-_{lm} \left( (\sigma_i)^{\hat{\alpha}}
_{\phantom{\alpha} \hat{\beta}}  \Omega^{\hat{\beta}}_{q_- lm} + 
\frac{1}{\kappa_-} \, L_i \Omega^{\hat{\alpha}}_{q_- lm} \right) \right\rbrace
\end{subeqnarray}
$v^{\hat{\alpha}}_{lm}$ is an arbitrary 2-component object for each value 
$\lbrace l,m \rbrace$ and $\xi^{(\pm)}_{lm}$ is a complex coefficient. 
We must apply the gauge-fixing condition to our expansion.
\begin{equation}
L_i \delta \Phi_i = \frac{1}{\sqrt{2}} \, L_i \left( a_i + i b_i \right) = 0
\end{equation}
Consider $v^{\hat{\alpha}}_{lm}$ to be,
\begin{eqnarray*}
v^{\hat{\alpha}}_{lm} = \left(\begin{array}{c} y_{lm} \\ z_{lm} \end{array}
\right)
\end{eqnarray*}
We can identify the fields $a_i$ and $b_i$,
\begin{eqnarray*}
\frac{1}{\sqrt{2}} \, a_i & = & \sum_{lm} \left\lbrace y_{lm} 
\frac{1}{\sqrt{l(l+1)}} \, i L_i Y_{lm} 
+ \sum_{+,-} \xi^{(\pm)}_{lm} \left(\sigma_i \Omega_{q_{\pm} lm} + 
\frac{1}{\kappa_{\pm}}L_i \Omega_{q_{\pm} lm} \right)^{\hat{1}} \right\rbrace 
\\
\frac{1}{\sqrt{2}} \, i b_i & = & \sum_{lm} \left\lbrace z_{lm} 
\frac{1}{\sqrt{l(l+1)}} \, i L_i Y_{lm}
+ \sum_{+,-} \xi^{(\pm)}_{lm} \left(\sigma_i \Omega_{q_{\pm} lm} + 
\frac{1}{\kappa_{\pm}}L_i \Omega_{q_{\pm} lm} \right)^{\hat{2}} \right\rbrace
\end{eqnarray*}
The T-spinor term is not constrained by the gauge-fixing as,
\begin{equation}
L_i \left(\sigma_i \Omega_{q_{\pm} lm} + \frac{1}{\kappa_{\pm}}L_i 
\Omega_{q_{\pm} lm} \right)^{\hat{\alpha}} = 0
\end{equation}
There is no constraint on $\xi^{(\pm)}_{lm}$.
Consider the remaining terms,
\begin{eqnarray*}
L_i \delta \Phi_i & = & L_i \left( \sum_{lm} y_{lm} \frac{1}{\sqrt{l(l+1)}} \, 
i L_i Y_{lm} + \sum_{lm} z_{lm} \frac{1}{\sqrt{l(l+1)}} \, i L_i Y_{lm} \right)
\\
& = & \sum_{lm} \left( y_{lm} + z_{lm} \right)\frac{1}{\sqrt{l(l+1)}} \, i L^2 
Y_{lm} = 0
\end{eqnarray*}
The gauge-fixing condition imposes the constraint $y_{lm} = - \, z_{lm}$ which 
allows us to choose,
\begin{eqnarray*}
v^{\hat{\alpha}}_{lm} = \frac{1}{\sqrt{2}} \, \left(\begin{array}{c} 1 \\ -1 
\end{array}\right) y_{lm} = v^{\hat{\alpha}} \, y_{lm}
\end{eqnarray*}
hence $v_{\hat{\alpha}}^{\dag} v^{\hat{\alpha}} = 1$. We can express the 
expansion $\mathcal{A}^{\hat{\alpha}}_i$ as,
\begin{equation}
\mathcal{A}^{\hat{\alpha}}_i = v^{\hat{\alpha}} \sum_{lm} y_{lm} 
\frac{1}{\sqrt{l(l+1)}} \, i L_i Y_{lm}
\end{equation}
The scalar field $a_i$ is real. This imposes a reality condition on the complex 
coefficient $y_{lm}$, $a_i = (a_i)^*$ implies,
\begin{eqnarray*}
\sum_{lm} y^{\phantom{\dag}}_{lm} i L_i Y^{\phantom{\dag}}_{lm} & = & \sum_{lm} 
y^*_{lm} i L_i Y^*_{lm} \\
& = & \sum_{lm} y^*_{lm} iL_i (-1)^m Y_{l, -m} \\
& = & \sum_{lm} y^*_{l, -m} (-1)^{-m} iL_i Y_{lm}
\end{eqnarray*}
hence,
\begin{equation}
y^*_{lm} = (-1)^{-m} y^{\phantom{\dag}}_{l, -m}
\end{equation}
\paragraph{}
The eigenvectors of the matrix are orthogonal, therefore the cross-terms are 
zero. (Suppressing our indices).
\begin{equation}\label{eq:scpot}
V = 8 \eta^4 \int d\Omega \left\lbrace \mathcal{A}^{\, \dag}_i \hat{O}_{ij} 
\mathcal{A}^{\, \phantom{\dag}}_j + \mathcal{P}^{\dag}_i \hat{O}_{ij} 
\mathcal{P}^{\phantom{\dag}}_j \right\rbrace
\end{equation}
The first term gives,
\begin{equation}
\int d\Omega \, \mathcal{A}^{\, \dag}_i \hat{O}_{ij} 
\mathcal{A}^{\, \phantom{\dag}}_j = R^4 \, \int d\Omega \sum_{lm,l'm'} 
y^{\dag}_{lm} \, y_{l'm'} \frac{1}{\sqrt{l(l+1)l'(l'+1)}} \, Y^{\dag}_{lm} 
\Delta^2_{S^2} Y_{l'm'}
\end{equation}
where the Laplacian on the 2-sphere is,
\begin{equation}
\Delta_{S^2} = \frac{1}{\sqrt{g}} \, \partial_{a} \left( g^{ab} \sqrt{g} \, 
\partial_b \right) = \frac{1}{R^2} \left( \textrm{cot} \, \theta \, 
\partial_{\theta} + \partial_{\theta} \partial_{\theta} + \textrm{csc}^2 \, \
theta \, \partial_{\phi} \partial_{\phi} \right)
\end{equation}
In section \ref{sec:sphere} we expanded vectors on the 2-sphere in the 
covariant and contravariant vector harmonic, $T_{lm \, a}$ and 
$T^{\phantom{lm} \, a}_{lm}$. The definition of these vector harmonics shows 
us how to define vector fields on the 2-sphere from spherical harmonics,
\begin{subeqnarray}
n_{\theta}(\theta, \phi) & = & R \, \sum_{lm} y_{lm} \frac{(-1)}{\sqrt{l(l+1)}}
\, \textrm{csc} \, \theta \, \partial_{\phi} Y_{lm}(\theta, \phi) \\
n_{\phi}(\theta, \phi) & = & R \, \sum_{lm} y_{lm} \frac{1}{\sqrt{l(l+1)}} \, 
\textrm{sin} \, \theta \, \partial_{\theta} Y_{lm}(\theta, \phi)
\end{subeqnarray}
Clearly the T-vector fields $n_a$ are real.
We can write,
\begin{eqnarray*}
R^2 \, \sum_{lm} y_{lm} \frac{1}{\sqrt{l(l+1)}} \, \Delta_{S^2} Y_{lm} & = & 
\sum_{lm} y_{lm} \frac{1}{\sqrt{l(l+1)}} \, \Big\lbrace \textrm{csc} \, \theta
\, \partial_{\theta}(\textrm{sin} \, \theta \, \partial_{\theta} Y_{lm}) \\
& &  \qquad + \textrm{csc} \, \theta \, \partial_{\phi}(\textrm{csc} \, \theta 
\, \partial_{\phi} Y_{lm}) \Big\rbrace \\
& = & \frac{1}{R} \, \big( \textrm{csc} \, \theta \, \partial_{\theta} n_{\phi}
- \textrm{csc} \, \theta \, \partial_{\phi} n_{\theta} \Big) \\
& = & \frac{1}{R} \, \textrm{csc} \, \theta \, \mathcal{F}_{\theta \phi}
\end{eqnarray*}
where $\mathcal{F}_{\theta \phi} = \partial_{\theta} n_{\phi} - \partial_{\phi}
n_{\theta}$. The first term of the scalar potential (\ref{eq:scpot}) is,
\begin{equation}
\begin{split}
8 \eta^4 \int d\Omega \, \mathcal{A}^{\, \dag}_i \hat{O}_{ij} 
\mathcal{A}^{\, \phantom{\dag}}_j & = 8 \eta^4 \frac{1}{R^2} \, \int d\Omega 
\, \textrm{csc}^2 \theta \mathcal{F}_{\theta \phi} \mathcal{F}_{\theta \phi} \\
& = 4 \eta^2 \int d\Omega \, \mathcal{F}_{ab} \mathcal{F}^{ab}
\end{split}
\end{equation}
as $\eta^2 \sim \frac{1}{R^2}$. 
Consider the second term of the scalar potential,
\begin{equation*}
\begin{split}
\int d\Omega & \, \mathcal{P}^{\dag}_i \hat{O}_{ij} 
\mathcal{P}^{\phantom{\dag}}_j = \int d\Omega \, \mathcal{P}^{\dag}_i 
\sum_{l'm'} \sum_{+,-} \xi^{(\pm)}_{l'm'} \, \hat{O}_{ij} \left( \sigma_j 
\Omega_{q'_{\pm} l'm'} + \frac{1}{\kappa'_{\pm}} L_j \Omega_{q'_{\pm}l'm'} 
\right) \\
& = R^2 \, \int d\Omega \, \mathcal{P}^{\dag}_i \sum_{l'm'} \sum_{+,-} 
\xi^{(\pm)}_{l'm'}\left( \sigma_i \kappa^2 \Omega_{q'_{\pm} l'm'} + 
\frac{1}{\kappa'_{\pm}} L_i \kappa^2 \Omega_{q'_{\pm} l'm'} \right) \\
& = R^2 \, \int d\Omega \sum_{lm,l'm'} \sum_{+,-} \left(\xi^{(\pm)}_{lm}
\right)
^{\dag} \xi^{(\pm)}_{l'm'} \Omega^{\dag}_{q_{\pm} lm} \left( 3\kappa^2 + 
\frac{1}{\kappa'_{\pm}} \, \sigma_i L_i \kappa^2 \right. \\
& \qquad \left. + \frac{1}{\kappa_{\pm}} \, \sigma_i L_i \kappa^2 + 
\frac{1}{\kappa_{\pm} \, \kappa'_{\pm}} L^2 \kappa^2 \right) 
\Omega_{q'_{\pm} l'm'}
\end{split}
\end{equation*}
(Note cross-terms involving ($q_+$, $q'_-$) and ($q_-$, $q'_+$) are zero due 
to the orthogonality of the spherical spinors).
Due to the orthogonality condition for the spherical spinors \eqref{eq:sphorth}
we can consider,
\begin{equation*}
\begin{split}
\int d\Omega \; \Omega^{\dag}_{q_{\pm}lm} & \left( 3 + \frac{1}{\kappa'_{\pm}}
\, \sigma_i L_i + \frac{1}{\kappa_{\pm}} \, \sigma_i L_i + 
\frac{1}{\kappa_{\pm} \, \kappa'_{\pm}} L^2 \right) \Omega_{q'_{\pm}l'm'} \\
& = \int d\Omega \; \Omega^{\dag}_{q_{\pm}lm} \left( 3 - 
\frac{2}{\kappa'_{\pm}} \, (\kappa'_{\pm} + 1) + \frac{1}{\kappa'^2_{\pm}} 
L^2 \right) \Omega_{q'_{\pm}l'm'} \\
& = \int d\Omega \; \Omega^{\dag}_{q_{\pm}lm} \left( 2 - 
\frac{1}{\kappa'_{\pm}} \right) \Omega_{q'_{\pm}l'm'}
\end{split}
\end{equation*}
Therefore if we redefine the complex coefficient,
\begin{eqnarray*}
\xi^{(\pm)}_{lm} \to \sqrt{2-\frac{1}{\kappa_{\pm}}} \, \xi^{(\pm)}_{lm}
\end{eqnarray*}
then,
\begin{equation}
\begin{split}
8 \eta^4 \, \int d\Omega \, \mathcal{P}^{\dag}_i \hat{O}_{ij} 
\mathcal{P}^{\phantom{\dag}}_j & = 8\eta^4 R^2 \, \int d\Omega \sum_{lm,l'm'}
\sum_{+,-} \left(\xi^{(\pm)}_{lm}\right)^{\dag} \xi^{(\pm)}_{l'm'} 
\Omega^{\dag}_{q_{\pm} lm} \kappa^2 \Omega_{q'_{\pm} l'm'} \\
& = 8\eta^2 \,\int d\Omega \, \xi^{\dag}_{\hat{\alpha}} \, 
(\kappa^2)^{\hat{\alpha}}_{\phantom{\alpha} \hat{\beta}} \xi^{\hat{\beta}}
\end{split}
\end{equation}
where,
\begin{equation}
\xi^{\hat{\alpha}}(\theta, \phi) = \sum_{lm} \sum_{+,-} \xi^{(\pm)}_{lm} 
\Omega^{\hat{\alpha}}_{q_{\pm}lm}(\theta, \phi)
\end{equation}
is a T-spinor field.
The scalar potential is,
\begin{eqnarray}
V = 8 \eta^2 \int d\Omega \left\lbrace \frac{1}{2} \mathcal{F}_{ab} 
\mathcal{F}^{ab} + \xi^{\dag}_{\hat{\alpha}} 
(\kappa^2)^{\hat{\alpha}}_{\phantom{\alpha} \hat{\beta}} \xi^{\hat{\beta}} 
\right\rbrace
\end{eqnarray}

We treat the fermionic part in a similar way.
\begin{eqnarray}
\mathcal{L}_{Fint} = \eta^4 \, \int d\Omega \left\lbrace i \bar{\Psi}_i 
\varepsilon_{ijk}L_k \Psi_j + 2i \bar{\Psi}_i L_i \Lambda - \bar{\Psi}_i 
\Psi_i \right\rbrace
\end{eqnarray}
Inspired by the SUSY tansformation of $\mathcal{Y}^{\hat{\alpha}}_i$ we define 
the following 2-component objects,
\begin{equation*}
\mathcal{X}^{\hat{\alpha}}_{i A} = \frac{1}{\sqrt{2}} \left( 
\begin{array}{c} \Psi_{i A} \\ 
i\left( \gamma_5 \right)^{\phantom{A} B}_A \Psi_{i B} \end{array} \right) 
\qquad \mathcal{Z}^{\hat{\alpha}}_A = \frac{1}{\sqrt{2}} \left( 
\begin{array}{c} \Lambda_{A} \\ 
i\left( \gamma_5 \right)^{\phantom{A} B}_A \Lambda_{B} \end{array} \right)
\end{equation*}
The interacting fermionic part is now,
\begin{equation}
\mathcal{L}_{Fint} = \eta^4 \, \int d\Omega \left\lbrace 
\bar{\mathcal{X}}_{i \hat{\alpha}}^{A} \big( \hat{\Delta}_{ij} 
\big)_{\hat{\beta} A}^{\hat{\alpha} B} \mathcal{X}_{j B}^{\hat{\beta}} 
+ 2i \, \bar{\mathcal{X}}_{i \hat{\alpha}}^{A} \left( 
\delta^{\hat{\alpha}}_{\phantom{\alpha} \hat{\beta}} 
\delta^{\phantom{A} B}_A  L_i \right) \mathcal{Z}_B^{\hat{\beta}} \right\rbrace
\end{equation}
with $\hat{\Delta}_{ij} = \delta^{\hat{\alpha}}_{\phantom{\alpha} \hat{\beta}}
\delta^{\phantom{A} B}_A  \left(i\varepsilon_{ijk} L_k - \delta_{ij} \right)$.
We expand the fermions of the chiral multiplets in the complete basis of 
eigenvectors of $\hat{O}_{ij}$,
\begin{equation}
\mathcal{X}^{\hat{\alpha}}_{i A} = 
\mathcal{B}^{\hat{\alpha}}_{i A} + \mathcal{R}^{\hat{\alpha}}_{i A}
\end{equation}
with,
\begin{subeqnarray}
\mathcal{B}^{\hat{\alpha}}_{i A} & = & \sum_{lm} u^{\hat{\alpha}}_{lm A} 
\frac{1}{\sqrt{l(l+1)}} \, iL_i \,Y_{lm}(\theta, \phi) \\
\mathcal{R}^{\hat{\alpha}}_{i A} & = & \sum_{lm} \sum_{+,-} 
\zeta^{(\pm)}_{lm A} \left( (\sigma_i)^{\hat{\alpha}}_{\phantom{\alpha} 
\hat{\beta}} \Omega^{\hat{\beta}}_{q_{\pm}lm}(\theta, \phi) + 
\frac{1}{\kappa_{\pm}} L_i \Omega^{\hat{\alpha}}_{q_{\pm}lm}(\theta, \phi)
\right)
\end{subeqnarray}
where $\zeta^{(\pm)}_{lm A}$ is a $SO(3,1)$ spinor coefficient. The coefficient
$u^{\hat{\alpha}}_{lm A}$ is a $SO(3,1)$ spinor for each $\lbrace l, m \rbrace$
and is a 2-component object like its bosonic counterpart $v^{\hat{\alpha}}_{lm}$.
\begin{equation*}
u^{\hat{\alpha}}_{lm A} = \frac{1}{\sqrt{2}} \left(\begin{array}{c} 
\delta^{\phantom{A} B}_A \\ i\left( \gamma_5 \right)^{\phantom{A} B}_A 
\end{array}\right) u_{lm B}
\end{equation*}
Consider the terms,
\begin{equation*}
\begin{split}
\eta^4 \, \int d\Omega \left\lbrace i \bar{\Psi}_i \varepsilon_{ijk} L_k \Psi_j - 
\bar{\Psi}_i \Psi_i \right\rbrace = \eta^4 \, \int d\Omega \Big\lbrace & 
\bar{\mathcal{B}}_i \hat{\Delta}_{ij} \mathcal{B}_j 
+ \bar{\mathcal{B}}_i \hat{\Delta}_{ij} \mathcal{R}_j \\
&  + \bar{\mathcal{R}}_i \hat{\Delta}_{ij} \mathcal{B}_j + \bar{\mathcal{R}}_i 
\hat{\Delta}_{ij} \mathcal{R}_j \Big\rbrace
\end{split}
\end{equation*}
The only non-zero terms are,
\begin{equation}
\eta^4 \, \int d\Omega \, \bar{\mathcal{R}}_i \hat{\Delta}_{ij}
\mathcal{R}_j = \eta^3 \, \int d\Omega \; \bar{\zeta}^A_{\hat{\alpha}}
(\theta, \phi) \, \kappa^{\hat{\alpha}}_{\phantom{\alpha} \hat{\beta}} \, 
\zeta^{\hat{\beta}}_A(\theta, \phi)
\end{equation}
where $\zeta(\theta, \phi)$ is a fermionic T-spinor,
\begin{equation}\label{eq:intfer}
\zeta^{\hat{\alpha}}_A (\theta, \phi) = \sum_{lm} \sum_{+,-} 
\zeta^{(\pm)}_{lm A} \Omega^{\hat{\alpha}}_{q_{\pm}lm}
\end{equation}
The remaining term of the interacting fermionic part is,
\begin{equation}
2\eta^2 \, \int d\Omega \, i\bar{\Psi}_i L_i \Lambda = 2\eta^2 \, \int d\Omega 
\left\lbrace i\bar{\mathcal{B}}_{i \hat{\alpha}}^{A} L_i 
\mathcal{Z}_A^{\hat{\alpha}} + i\bar{\mathcal{R}}_{i \hat{\alpha}}^{A} L_i 
\mathcal{Z}_A^{\hat{\alpha}}\right\rbrace
\end{equation}
which contains the gaugino $\Lambda$. The 2-component object $\varLambda_A$ is 
expanded in spherical harmonics,
\begin{equation}
\mathcal{Z}^{\hat{\alpha}}_A(\theta,\phi) = \sum_{lm} 
\mathcal{Z}^{\hat{\alpha}}_{lm A} Y_{lm}(\theta, \phi)
\end{equation}
where $\mathcal{Z}^{\hat{\alpha}}_{lm A}$ is a $SO(3,1)$ spinor coefficient and
an arbitrary 2-component object like $u^{\hat{\alpha}}_{lm A}$.
\begin{equation*}
\mathcal{Z}^{\hat{\alpha}}_{lm A} = \frac{1}{\sqrt{2}} \left(\begin{array}{c} \
delta^{\phantom{A} B}_A \\ i\left( \gamma_5 \right)^{\phantom{A} B}_A 
\end{array}\right) \Lambda_{lm B}
\end{equation*}
The second term of \eqref{eq:intfer} vanishes, but the first term is,
\begin{equation}
\int d\Omega \; i\bar{\mathcal{B}}^{\hat{\alpha}}_{i A}  L_i 
\mathcal{Z}^A_{\hat{\alpha}} = R \, \int d\Omega \, \frac{1}{\sqrt{g}} \, 
\bar{\mathcal{G}}^A_{\theta \phi} \Lambda_A
\end{equation}
where $\mathcal{G}_{ab} = \partial_a g_b - \partial_b g_a$ is the `field 
tensor' of the fermionic T-vector $g_a(\theta, \phi)$ in analogy with the 
bosons,
\begin{subeqnarray}
g_{\theta A}(\theta, \phi) & = & R \, \sum_{lm} u_{lm A} \frac{(-1)}{\sqrt{l(l+1)}}
\, \textrm{csc} \, \theta \, \partial_{\phi} Y_{lm}(\theta, \phi) \\
g_{\phi A}(\theta, \phi) & = & R \, \sum_{lm} u_{lm A} \frac{1}{\sqrt{l(l+1)}} \,
 \textrm{sin} \, \theta \, \partial_{\theta} Y_{lm}(\theta, \phi) \\
\frac{1}{\sqrt{g}} \, \mathcal{G}_{\theta \phi A} & = & R \, \sum_{lm} u_{lm A}
 \frac{1}{\sqrt{l(l+1)}} \, \Delta_{S^2} Y_{lm}
(\theta, \phi)
\end{subeqnarray}
and,
\begin{equation*}
\Lambda_A(\theta,\phi) = \sum_{lm} \Lambda_{lm A} 
Y_{lm}(\theta, \phi)
\end{equation*}
Therefore,
\begin{equation}
2\eta^4 \int d\Omega \, i\bar{\mathcal{B}}_i L_i \mathcal{Z} = \eta^3 \, 
\int d\Omega \, \frac{1}{\sqrt{g}} \, \varepsilon^{ab} \bar{\mathcal{G}}_{ab} 
\Lambda
\end{equation}
where $\varepsilon^{\theta \phi} = 1$.
The fermionic part has become,
\begin{equation}
\mathcal{L}_{Fint} = \eta^3 \, \int d\Omega \Big\lbrace 
\bar{\zeta}_{\hat{\alpha}} \, \kappa^{\hat{\alpha}}_{\phantom{\alpha} 
\hat{\beta}} \, \zeta^{\hat{\beta}} + \frac{1}{\sqrt{g}} \, \varepsilon^{ab} 
\bar{\mathcal{G}}_{ab} \Lambda \Big\rbrace
\end{equation}

Consider the terms originating from the $\mathcal{N}=1^*$ bosonic kinetic 
terms,
\begin{equation}\label{eq:boskin}
\begin{split}
\mathcal{L}_{Bkin} = \frac{1}{g^2_{ym}} \frac{N}{4 \pi R^2} \, \eta^2 \int &
d\Omega \Big\lbrace -\frac{1}{4} F_{\mu \nu} F^{\mu \nu} - 2 \partial_{\mu} 
(\delta \Phi^{\dag}_i) \partial^{\mu} (\delta \Phi^{\phantom{\dag}}_i) \qquad 
\\
& + 2\eta^2(L_i A_{\mu})(L_i A^{\mu}) \Big\rbrace
\end{split}
\end{equation}
The gauge boson is expanded in spherical harmonics,
\begin{equation}
A_{\mu}(\theta, \phi) = \sum_{lm} A_{(\mu)lm} Y_{lm}(\theta, \phi)
\end{equation}
where $A_{(\mu)lm}$ is a complex coefficient.
The first term in \eqref{eq:boskin} is trivial. The third term is clearly,
\begin{eqnarray}
2\eta^2 \int d\Omega (L_i A_{\mu})(L_i A^{\mu}) = 2\int d\Omega \; A_{\mu} 
\Delta_{S^2} A^{\mu}
\end{eqnarray}
The second term is,
\begin{eqnarray*}
2\eta^2 \, \int d\Omega \, \partial_{\mu} (\delta \Phi^{\dag}_i) \partial^{\mu}
(\delta  \Phi^{\phantom{\dag}}_i) 
& = & 2\eta^2 \, \int d\Omega \; \partial_{\mu} \mathcal{Y}^{\dag}_i 
\partial^{\mu} \mathcal{Y}^{\phantom{\dag}}_i \\
& = & 2\eta^2 \, \int d\Omega \; \Big( \partial_{\mu} \mathcal{A}^{\dag}_i 
\partial^{\mu} \mathcal{A}^{\phantom{\dag}}_i + \partial_{\mu} 
\mathcal{P}^{\dag}_i \partial^{\mu} \mathcal{P}^{\phantom{\dag}}_i \Big)
\end{eqnarray*}
We ignore the cross-terms due to the orthogonality of the eigenvectors,
\begin{equation*}
\begin{split}
2\eta^2 \, \int d\Omega \, \partial_{\mu} & \mathcal{A}^{\dag}_i \partial^{\mu}
\mathcal{A}^{\phantom{\dag}}_i \\
& = 2\eta^2 \, \int d\Omega \sum_{lm} \sum_{l'm'} \partial_{\mu} y^{\dag}_{lm} 
\, \partial^{\mu} y_{l'm'} 
\frac{1}{\sqrt{l(l+1)l'(l'+1)}} \, Y^{\dag}_{lm} L^2 Y_{l'm'}
\end{split}
\end{equation*}
We can use equation (\ref{eq:killing}) to re-express,
\begin{eqnarray*}
\sum_{lm} y_{lm} \frac{1}{\sqrt{l(l+1)}} \, iL_i \, Y_{lm} & = & \sum_{lm} 
y_{lm} \frac{1}{\sqrt{l(l+1)}} \, k^a_i \partial_a Y_{lm} = \frac{1}{R} \,
\frac{1}{\sqrt{g}} \, k^{a}_i g_{ab} \, \varepsilon^{bc} n_{c}
\end{eqnarray*}
Therefore,
\begin{equation}
\begin{split}
2\eta^2 \, \int d\Omega \, \partial_{\mu} \mathcal{A}^{\dag}_i \partial^{\mu} 
\mathcal{A}^{\phantom{\dag}}_i & = 2\eta^2 \, \frac{1}{R^2} \, \int d\Omega \, 
\partial_{\mu} \left( \frac{1}{\sqrt{g}} \, k^{a}_i g_{ab} \, \varepsilon^{bc} 
n_{c} \right) \partial^{\mu} \left( \frac{1}{\sqrt{g}} \, k^{d}_i g_{de} \, 
\varepsilon^{ef} n_{f} \right) \\
& = 2\eta^2 \, \frac{1}{R^2} \, \int d\Omega \, \frac{1}{g} \, \varepsilon^{bc}
g_{be} \, \varepsilon^{ef} \partial_{\mu} n_c \, \partial^{\mu} n_f \\
& = 2\eta^2 \, \int d\Omega \, \partial_{\mu} n_{a} \partial^{\mu} n^{a}
\end{split}
\end{equation}
The final term of the bosonic kinetic term is,
\begin{equation}
\begin{split}
\int d\Omega \, \partial_{\mu} \mathcal{P}^{\dag}_i \partial^{\mu} 
\mathcal{P}^{\phantom{\dag}}_i = & \int d\Omega \sum_{lm,l'm'} \sum_{+,-} 
\partial_{\mu}(\xi^{(\pm)}_{lm})^{\dag} \partial^{\mu} \xi^{(\pm)}_{l'm'} 
\,  \left( \Omega^{\dag}_{q_{\pm}lm} \sigma_i + \frac{1}{\kappa_{\pm}} 
\Omega^{\dag}_{q_{\pm}lm} L_i \right) \\
& \qquad \times \left( \sigma_i \Omega^{\phantom{\dag}}_{q'_{\pm}l'm'} + 
\frac{1}{\kappa'_{\pm}} L_i \Omega^{\phantom{\dag}}_{q'_{\pm}l'm'} \right) \\
\to & \int d\Omega \, \partial_{\mu} \xi^{\dag}_{\hat{\alpha}}(\theta, \phi) 
\partial^{\mu} \xi^{\hat{\alpha}}(\theta, \phi)
\end{split}
\end{equation}

The kinetic term of the gauginos is trivial. The kinetic term of the chiral 
spinors is,
\begin{equation}
\begin{split}
\eta^3 \,\int d\Omega \, \bar{\Psi}^A_i \left(\gamma^{\mu}
\right)^{\phantom{A} B}_A  \partial_{\mu} \Psi_{i B} & = 
\int d\Omega \, \bar{\mathcal{X}}^A_{i \hat{\alpha}} \left(\gamma^{\mu}
\right)^{\phantom{A} B}_A  \partial_{\mu} \mathcal{X}^{\hat{\alpha}}_{i B} \\
& = \eta^3 \,\int d\Omega \Big\lbrace \bar{\mathcal{B}}^A_{i \hat{\alpha}} 
\left(\gamma^{\mu}\right)^{\phantom{A} B}_A  \partial_{\mu} 
\mathcal{B}^{\hat{\alpha}}_{i B} + \bar{\mathcal{R}}^A_{i \hat{\alpha}} 
\left(\gamma^{\mu}\right)^{\phantom{A} B}_A  \partial_{\mu} 
\mathcal{R}^{\hat{\alpha}}_{i B} \Big\rbrace 
\end{split}
\end{equation}
Due to the orthogonality of the eigenvectors we ignore the cross-terms. 
Consider,
\begin{equation*}
\begin{split}
\eta^3 \, \int d\Omega \, \bar{\mathcal{B}}^A_{i \hat{\alpha}} \left(
\gamma^{\mu}\right)^{\phantom{A} B}_A  \partial_{\mu} 
\mathcal{B}^{\hat{\alpha}}_{i B} = \eta^3 \, \int d\Omega \, & \sum_{lm} 
\sum_{l'm'} \left( \bar{u}^A_{lm} 
\left(\gamma^{\mu}\right)^{\phantom{A} B}_A \partial_{\mu} u_{l'm' B} \right) \\
& \times \frac{1}{\sqrt{l(l+1)l'(l'+1)}} \, Y^{\dag}_{lm} L^2 Y_{l'm'}
\end{split}
\end{equation*}
Now in analogy with the bosons,
\begin{equation}
\sum_{lm} u_{lm A} \frac{1}{\sqrt{l(l+1)}} \, iL_i Y_{lm} = \frac{1}{R} \, 
\frac{1}{\sqrt{g}} \, k^a_i g_{ab} \varepsilon^{bc} g_{c A}
\end{equation}
Therefore,
\begin{equation}
\begin{split}
\eta^3 \int d\Omega \, \bar{\mathcal{B}}^A_{i \hat{\alpha}} \left(\gamma^{\mu}
\right)^{\phantom{A} B}_A  \partial_{\mu} \mathcal{B}^{\hat{\alpha}}_{i B} & =
\eta^3 \int d\Omega \, \frac{1}{R^2} \, \frac{1}{g} \, k^a_i g_{ab} 
\varepsilon^{bc} k^d_i g_{de} \varepsilon^{ef} \bar{g}^A_{c} \left(\gamma^{\mu}
\right)^{\phantom{A} B}_A \partial_{\mu} g_{f B} \\
& = \eta^3 \int d\Omega  \, \bar{g}^A_a 
\left(\gamma^{\mu}\right)^{\phantom{A} B}_A \partial_{\mu} g_B^a
\end{split}
\end{equation}
Consider the final term,
\begin{equation}
\begin{split}
\eta^3 \int d\Omega \, \bar{\mathcal{R}}^A_{i \hat{\alpha}} & 
\left(\gamma^{\mu}\right)^{\phantom{A} B}_A  \partial_{\mu} 
\mathcal{R}^{\hat{\alpha}}_{i B} \\
& = \eta^3 \int d\Omega \, \sum_{lm} \sum_{l'm'} \sum_{+,-} 
\left( \bar{\zeta}^{(\pm)A}_{lm} \left(\gamma^{\mu}\right)^{\phantom{A} B}_A 
\partial_{\mu} \zeta^{(\pm)}_{l'm' B}\right) \\
& \qquad \times \left( \Omega^{\dag}_{q_{\pm}lm} 
\sigma_i + \frac{1}{\kappa_{\pm}} \Omega^{\dag}_{q_{\pm}lm} L_i 
\right)_{\hat{\alpha}} \left( \sigma_i 
\Omega^{\phantom{\dag}}_{q'_{\pm}l'm'} + \frac{1}{\kappa'_{\pm}} L_i 
\Omega^{\phantom{\dag}}_{q'_{\pm}l'm'} \right)^{\hat{\alpha}}
\\
& \to \eta^3 \int d\Omega \, \bar{\zeta}^A_{\hat{\alpha}} 
\left(\gamma^{\mu}\right)^{\phantom{A} B}_A \partial_{\mu} 
\zeta_B^{\hat{\alpha}}
\end{split}
\end{equation}

Summarizing the above calculations, the resulting six-dimensional effective 
action (contracting all spinor indices),
\begin{equation}\label{eq:MNaction}
\begin{split}
\mathcal{S} = \frac{1}{g^2_{ym}} \frac{N}{4 \pi} & \, \eta^2 
\int d^4 x \, \int d\Omega \Big\lbrace - \frac{1}{4} F_{\mu \nu} F^{\mu \nu} - 
\frac{i}{2} \, \eta \bar{\Lambda} \gamma^{\mu} \partial_{\mu} \Lambda - 
\frac{i}{2} \, \eta \bar{g}_a \gamma^{\mu} \partial_{\mu} g^a \\
& - \frac{i}{2} \, \eta \bar{\zeta} \gamma^{\mu} \partial_{\mu} \zeta - 2
\partial_{\mu} n_{a} \partial^{\mu} n^{a} -2\partial_{\mu} \xi^{\dag} 
\partial^{\mu} \xi + 2 A_{\mu} \Delta_{S^2} A^{\mu} \\
&- 4 \mathcal{F}_{ab} \mathcal{F}^{ab}  - 8\xi^{\, \dag} \kappa^2
\xi + \eta \, \bar{\zeta} \, \kappa \, \zeta +  \frac{1}{\sqrt{g}} \, \eta 
\, \varepsilon^{ab}  \bar{\mathcal{G}}_{ab} \Lambda 
\Big\rbrace
\end{split}
\end{equation}
The T-spinors are all expressed in terms of the cartesian basis of spherical 
spinors, however in our calculation of the MN spectrum we used spherical 
spinors in the spherical basis, therefore we will express the action in this 
basis.
\begin{equation}
\xi = V^{\dag} \Xi \qquad \qquad \zeta = V^{\dag} \Upsilon
\end{equation}
Which gives the action,
\begin{equation}\label{eq:MNaction1}
\begin{split}
\mathcal{S} = \frac{1}{g^2_{ym}} & \frac{N}{4 \pi R^2} \, \eta^2 
\int d^4 x \, \int R^2 d\Omega \Big\lbrace - \frac{1}{4} F_{\mu \nu} 
F^{\mu \nu} - \frac{i}{2} \, \eta \, \bar{\Lambda} \gamma^{\mu} \partial_{\mu} 
\Lambda - \frac{i}{2} \, \eta \, \bar{g}_a \gamma^{\mu} \partial_{\mu} g^a \\
& - \frac{i}{2} \, \eta \, \bar{\Upsilon} \gamma^{\mu} \partial_{\mu} 
\Upsilon  - 2\partial_{\mu} n_{a} \partial^{\mu} n^{a} -2\partial_{\mu} 
\Xi^{\dag} \partial^{\mu} \Xi + 2 A_{\mu} \Delta_{S^2} A^{\mu} 
- 4 \mathcal{F}_{ab} \mathcal{F}^{ab} \\
& - 8 \, \Xi^{\dag} (-i \hat{\nabla}_{S^2})^2 \, \Xi + \eta \, \bar{\Upsilon}
\hat{\gamma}_3 (-i \hat{\nabla}_{S^2}) \Upsilon + \frac{1}{\sqrt{g}} \, \eta 
\, \varepsilon^{ab} \bar{\mathcal{G}}_{ab} \Lambda 
\Big\rbrace
\end{split}
\end{equation}

The above action has the canonical 
form of a six-dimensional theory and the calculation of Sections 5
implies that it has the 
correct spectrum. The action contains no interaction terms as one would expect 
in a $U(1)$ gauge theory with adjoint matter and there are also no explicit 
mass terms. The six-dimensional theory has intrinsically massless fields 
propagating along the flat four dimensional spacetime and along the compact 
2-manifold. It is the compactness of the 2-manifold which gives mass to the 
theory. This is to be expected as the MN compactification is a twisted 
compactification of the massless $\mathcal{N}=(1,1)$ SUSY Yang-Mills theory.
The fields are summarized below.
\begin{center}
\begin{tabular}{ccc}
\hline
Fields & T-Spin & Spin \\
\hline
$A_{\mu}$ & T-scalar & \\
$\Xi$ & T-spinor & Bosons\\
$n_a$ & T-vector & \\
\hline
$\Lambda$ & T-scalar & \\
$\Upsilon$ & T-spinor & Fermions \\
$g_a$ & T-vector & \\
\hline
\end{tabular}
\end{center}

We can see how the twisted compactification of Maldacena and N\'u\~nez is 
realized in the classical action. The bosonic T-scalars, fermionic T-spinors 
and bosonic T-vectors are all realized as in an untwisted field theory. The 
kinetic term of the bosonic T-spinors on the 2-sphere is realized as the square
of the Dirac operator. This makes perfect sense. The kinetic term of a boson is
quadratic in derivatives whilst a fermion is linear in derivatives. It is not 
possible to remove a derivative from the action, therefore the bosonic 
T-spinors must be quadratic in derivatives. In a standard field theory the 
spinors have a Dirac operator, linear in derivatives, so it is sensible for 
the kinetic term of the bosonic T-spinor to be the square of the Dirac operator
on the sphere. We find the fermionic T-scalars and fermionic T-vectors have a 
coupled kinetic term. We can think of this term as a combined square root of a 
scalar Laplacian and a Maxwell term. This makes sense for the same reason as 
the bosonic T-spinor. The fields are fermions and hence linear in derivatives, 
so the kinetic term for the fermionic T-scalars and T-vectors must be a 
``square root'' of the canonical kinetic term for scalars and vectors, 
respectively. 
The origin of the ``coupled kinetic term'' can be seen in the 
$\mathcal{N}=(1,1)$ action \eqref{eq:n11act}. The fields 
$\lambda^A_{\underline{\alpha}}$ are in the ${\bf 4}$ representation of $SU(4)$
. It is this $\mathcal{N}=(1,1)$ spinor which forms the fermionic T-scalars and
fermionic T-vectors of the Maldacena-N\'u\~nez compactification. Therefore the 
term,
\begin{eqnarray*}
\lambda^{A \underline{\alpha}} \bar{\Sigma}^i_{AB} \partial_i 
\lambda^{B}_{\underline{\alpha}}
\end{eqnarray*}
from the $\mathcal{N}=(1,1)$ action will upon twisted compactification give 
rise to the term coupling the fermionic T-scalars and fermionic T-vectors. 
Futhermore, the Kaluza-Klein spectrum of the MN compactifcation suggests a 
coupling between the fermionic T-scalars and T-vectors. In order to organise 
the MN fields into $\mathcal{N}=1$ multiplets we had to combine the T-scalars 
with the T-vectors to obtain massive vector multiplets.

Finally we can also see that the effective action calculated above is 
identical to the bosonic action (\ref{result}) derived from the 
$\mathcal{N}=(1,1)$ SUSY Yang-Mills action. In particular, 
with appropriate rescaling of the fields\footnote{This including the absorbing 
of the mass parameter $\eta$ to return the fields to the correct mass dimension.}, 
fixing the gauge $\partial_a n^a = 0$ and the integration by parts of the 
kinetic term of the $U(1)$ gauge field, (\ref{result}) 
is identical to the bosonic part of (\ref{eq:MNaction}), with six-dimensional 
coupling $g^2_6 = 4\pi R^2 g^2_{ym}/N$. 
As the $\mathcal{N}=(1,1)$ SUSY 
Yang-Mills theory is supersymmetric this is sufficient evidence to show that 
the Maldacena-N\'u\~nez compactification is equivalent to the Higgsed 
$\mathcal{N}=1^*$ SUSY Yang-Mills theory in the limit $N \to \infty$.

\paragraph{}
The work of ND is supported by a PPARC senior fellowship. RPA would like to 
thank Tim Hollowood and Carlos N\'u\~nez for their help, support and 
suggestions.

\appendix

\section{Clifford Algebras}\label{app:1}

Our notation generally follows that of Wess and Bagger \cite{bib:WB}, with 
metric convention $(-,+,+,+,\dots)$. For the four-dimensional theories we 
use the $SO(3,1)$ Clifford algebra of Wess and Bagger,
\begin{equation}
\gamma^{\mu} = \left( \begin{array}{cc} 0 & \sigma^{\mu} \\ 
\bar{\sigma}^{\mu} & 0 \end{array}\right)
\end{equation}
The $SO(3,1)$ chirality matrix is,
\begin{equation}
\gamma^5 = \gamma^0 \gamma^1 \gamma^2 \gamma^3 = \left( \begin{array}{cc} 
-i & 0 \\ 0 & i \end{array}\right)
\end{equation}

We derive the $\mathcal{N}=(1,1)$ SUSY Yang-Mills theory in a six-dimensional 
spacetime from the $\mathcal{N}=1$ SUSY Yang-Mills theory in a ten-dimensional 
spacetime, via Trivial Dimensional Reduction. The action for $\mathcal{N}=1$ SUSY Yang-Mills
in ten spacetime dimensions is,
\begin{eqnarray}
\mathcal{S} = \frac{1}{g^2} \int d^{10} x \left( -\frac{1}{4}F_{MN}F^{MN} - 
\frac{i}{2} \bar{\Psi} \Gamma^M \partial_{M} \Psi \right)
\end{eqnarray}
$F_{MN} = \partial_M A_N - \partial_N A_M$, where $A_M$ is a ten-dimensional 
gauge field and $\Psi$ is a 32-component spinor of $SO(9,1)$, 
$M = 0,1,\dots, 9$.
Trivial dimensional reduction assumes the ten dimensional fields in the 
above action are dependent on only the first six dimensions $x^i$, where 
$i = 0,1,\dots, 5$.
\begin{eqnarray}
\nonumber F_{MN} F^{MN} & = & F_{ij} F^{ij} + 2 \, \partial_i \phi_m \, 
\partial^i \phi^m
\end{eqnarray}
where $\phi_m = A_m$, $m = 1,2,3,4$.
To calculate the spinor contribution we must decompose the Majorana-Weyl 
spinor of $SO(9,1)$ in to a representation of $SO(5,1)\times SO(4)$.
\begin{equation}
{\bf 16} \to ( {\bf 4}, {\bf 2}, {\bf 1}) \oplus ( {\bf \bar{4}}, {\bf 1}, 
{\bf 2})
\end{equation}
This decomposition is obtained by the following Clifford algebra decomposition
 of $SO(9,1)$.
\begin{equation}
\Gamma^M = \left\lbrace \tilde{\Gamma}^i \otimes \tilde{\gamma}^5, \mathbb{1}_8
 \otimes \tilde{\gamma}^m \right\rbrace
\end{equation}
where $\tilde{\Gamma}^i$ is the $SO(5,1)$ Clifford algebra and 
$\tilde{\gamma}^m$ is the $SO(4)$ Clifford algebra. The algebra for $SO(5,1)$ 
follows from \cite{bib:DHKM},
\begin{equation}
\tilde{\Gamma}^i = \left( \begin{array}{cc} 0 & \Sigma^i \\ \bar{\Sigma}^i & 0 \end{array}\right)
\end{equation}
where,
\begin{subeqnarray}
\Sigma^i & = & \left( -i\eta^3, i\bar{\eta}^3, \eta^2, i\bar{\eta}^2, \eta^1, 
i\bar{\eta}^1 \right) \\
\bar{\Sigma}^i & = & \left( i\eta^3, i\bar{\eta}^3, -\eta^2, i\bar{\eta}^2, 
-\eta^1, i\bar{\eta}^1 \right)
\end{subeqnarray}
where $\eta^c$ and $\bar{\eta}^c$ are the t'Hooft eta symbols.
The algebra for $SO(4)$ is \cite{bib:DHKM},
\begin{equation}
\tilde{\gamma}^m = \left( \begin{array}{cc} 0 & \tau^m \\ \bar{\tau}^m & 0 
\end{array}\right)
\end{equation}
where $\tau_m = (\vec{\sigma}, -i\mathbb{1})$ and $\bar{\tau}_m = (\vec{\sigma}
, i\mathbb{1})$.

From the chirality condition the spinor $\Psi$ decomposes into,
\begin{equation}
\Psi = \left(\begin{array}{c} 1 \\ 0 \end{array} \right) \otimes 
\left(\begin{array}{c} 1 \\ 0 \end{array} \right) 
\lambda^A_{\underline{\alpha}} + \left(\begin{array}{c} 0 \\ 1 \end{array} 
\right) \otimes \left(\begin{array}{c} 0 \\ 1 \end{array} \right) \bar{\lambda}^{\underline{\dot{\alpha}}}_A
\end{equation}
where $A= 1,2,3,4$ is an $SO(5,1)$ spinor index and $\underline{\alpha}, \underline{\dot{\alpha}} = 1,2$ are $SU(2)_A \times SU(2)_B \sim SO(4)$ spinor
indices. From the Majorana condition the Weyl spinors have the following 
hermitian conjugates.
\begin{subeqnarray}
\left( \lambda^A_{\underline{\alpha}} \right)^{\dag} & = & \bar{\Sigma}^0_{AB} 
\lambda^{B \underline{\alpha}} \\
\left(\bar{\lambda}^{\underline{\dot{\alpha}}}_A\right)^{\dag} & = & 
\Sigma^{0AB} \bar{\lambda}_{B \underline{\dot{\alpha}}}
\end{subeqnarray}
The fermionic term is,
\begin{eqnarray}
i \bar{\Psi} \Gamma^M \partial_M \Psi & = & i \lambda^{A \underline{\alpha}} \bar{\Sigma}^i_{AB} \partial_i \lambda^{B}_{\underline{\alpha}} + 
i \bar{\lambda}_{A \underline{\dot{\alpha}}} \Sigma^{i AB} \partial_i \bar{\lambda}^{\underline{\dot{\alpha}}}_B
\end{eqnarray}
and the full $\mathcal{N}=(1,1)$ SUSY Yang-Mills actions is,
\begin{eqnarray}\label{eq:n11act}
\nonumber \mathcal{S} & = & \frac{1}{g^2_6} \int d^{6} x \left( -\frac{1}{4}
F_{ij} F^{ij} - \frac{1}{2} \, \partial_i \phi_m \, \partial^i \phi^m \right) 
\\
&& \qquad \qquad \qquad \left. - \frac{i}{2} \lambda^{A \underline{\alpha}} 
\bar{\Sigma}^i_{AB} \partial_i \lambda^{B}_{\underline{\alpha}} - \frac{i}{2} 
\bar{\lambda}_{A \underline{\dot{\alpha}}} \Sigma^{i AB} \partial_i \bar{\lambda}^{\underline{\dot{\alpha}}}_B \right)
\end{eqnarray}

The $SO(4)$ Clifford algebra has the following identities,
\begin{subeqnarray}
\textrm{Tr} \, \tau^m \bar{\tau}^n & = & 2 \, \delta^{mn} \\
(\tau^m)_{\underline{\alpha} \underline{\dot{\alpha}}} 
(\bar{\tau}_m)^{\underline{\dot{\beta}} \underline{\beta}} & = & 2 \,  
\delta^{\phantom{\alpha} \underline{\beta}}_{\underline{\alpha}} 
\delta^{\phantom{\alpha} \underline{\dot{\beta}}}_{\underline{\dot{\alpha}}} \\
(\bar{\tau}^m)_{\underline{\alpha} \underline{\dot{\alpha}}} (\bar{\tau}_m)^{\underline{\dot{\beta}} \underline{\beta}} & = & -2 \, \varepsilon^{\underline{\alpha} \underline{\beta}} 
\varepsilon^{\underline{\dot{\alpha}} \underline{\dot{\beta}}}
\end{subeqnarray}
The $SO(4)$ bispinor has the hermitian conjugate,
\begin{eqnarray*}
\left( v^{\phantom{\alpha} \underline{\dot{\alpha}}}_{\underline{\alpha}} 
\right)^{\dag} & = & (\bar{\lambda}^{\underline{\dot{\alpha}}}_A)^{\dag} (\lambda^A_{\underline{\alpha}})^{\dag} \\
& = & \Sigma^{0 AB} \bar{\lambda}_{B \underline{\dot{\alpha}}} 
\bar{\Sigma}^0_{AC} \lambda^{C \underline{\alpha}}\\
& = & \bar{\lambda}_{A \underline{\dot{\alpha}}} \lambda^{A \underline{\alpha}}
\\
& = & -\lambda^{A \underline{\alpha}} 
\bar{\lambda}_{A \underline{\dot{\alpha}}}\\
& = & -v^{\underline{\alpha}}_{\phantom{\alpha} \underline{\dot{\alpha}}}
\end{eqnarray*}

\end{document}